\documentclass[reprint,nofootinbib]{revtex4-2}
\usepackage{amsmath}
\usepackage{amssymb}
\usepackage{xcolor}
\usepackage{autobreak}
\usepackage{float}%H
\usepackage{graphicx}
\graphicspath{{figures/}}
\usepackage[colorlinks=true,linkcolor=red,urlcolor=blue,citecolor=blue,bookmarksopen=true,bookmarks=true]{hyperref}

\begin{document}

\title{Probing vector gravitational atoms with eccentric intermediate mass-ratio inspirals}

\author{Yan Cao$^{a}$}
\email{yancao@smail.nju.edu.cn}
\author{Ya-Ze Cheng$^{b}$}
\author{Gen-Liang Li$^{c}$}
\author{Yong Tang$^{b,d,e}$}
\email{tangy@ucas.ac.cn}
\affiliation{\begin{footnotesize}
		${}^a$School of Physics, Nanjing University, Nanjing 210093, China\\
		${}^b$School of Astronomy and Space Sciences, University of Chinese Academy of Sciences (UCAS), Beijing 100049, China\\
        ${}^c$New Engineering Industry College, Putian University, Putian 351100, China\\
		${}^d$School of Fundamental Physics and Mathematical Sciences, \\
		Hangzhou Institute for Advanced Study, UCAS, Hangzhou 310024, China \\
		${}^e$International Center for Theoretical Physics Asia-Pacific, Beijing/Hangzhou, China
\end{footnotesize}}

\date{\today}

\begin{abstract}
Ultralight bosons, proposed as candidates for dark matter (DM), are predicted by various new physics models. In the presence of bosons with suitable masses, superradiant (SR) instability can naturally transform a spinning black hole (BH) into a gravitational atom (GA). Here we study the dynamics of intermediate mass-ratio inspirals (IMRIs) around a GA formed by ultralight vector field saturated in its SR ground state. We employ a perturbative model at the leading Newtonian order to consistently account for both the conservative effect of cloud gravity and the dissipative effect of cloud ionization. We find the cloud can make a sizable negative contribution to the secular periastron precession at binary separations comparable to the gravitational Bohr radius. Meanwhile, the backreaction of ionization could significantly accelerate the process of orbital decay and circularization. Considering reasonably small vector boson masses, we examine the adiabatic orbital evolution and gravitational waveforms of eccentric inspirals. The results indicate that vector GAs might be detectable through observations of low-frequency IMRIs by the future space-based gravitational-wave detectors, such as LISA and Taiji.
\end{abstract}

\maketitle

\newpage
\section{Introduction}
New bosons with masses below $30\,\text{eV}$ are predicted by various physics theories beyond the standard model, and are among the leading dark matter (DM) candidates \cite{Peccei.Quinn, Wilczek.1978, Preskill:1982cy, Abbott:1982af, Dine:1982ah, Holdom:1985ag, Svrcek:2006yi, Arvanitaki_2010, Hui:2016ltb, Hui:2021tkt, Ferreira_2021, Matos:2023usa}. Even if they constitute all the DM, experimental search of these ultralight bosons will be particularly challenging if they interact with normal matter only through gravity \cite{Aoki:2016kwl, Kim:2023pkx,Kim:2023kyy, Yu:2024enm, Kim:2024xcr,Dror:2024con,An:2024axz, Wang:2023phr,Blas:2024duy}. On the other hand, ultralight bosons may form dense classical field condensates throughout our Universe, which leads to the intriguing possibility to probe them through the observations of purely gravitational processes.

Precise measurements of electromagnetic or gravitational waves (GWs) emitted by the compact binaries can be used to probe these ultralight bosonic fields, e.g., through their imprints on the binary's orbital motion \cite{Wong:2019yoc, Wong:2020qom, Blas:2019hxz, Annulli:2020lyc, Vicente:2022ivh, Pombo:2023ody, Traykova:2023qyv, PhysRevD.107.024035, PhysRevLett.132.211401,Brax:2024yqh, kim2024gravitationalwaveduetresonating,2024PhLB..85638908K,Bromley:2023yfi}. One interesting possibility is that dense bosonic fields are bound to the bodies in the binary. Stellar-mass compact objects orbiting a massive central black hole (BH) (forming an extreme mass-ratio or intermediate mass-ratio binary system, for central BH mass larger than $10^6M_\odot$ or within $200\sim 10^5M_\odot$, respectively) would be the ideal targets to probe the bosonic field structure around the central BH, since a small companion can respond more sensitively to such BH environment but without destroying it \cite{Kocsis:2011dr, Eda:2013gg, Barausse:2014tra, Yue:2018vtk, Yue:2019ozq, Bertone:2019irm, Li:2021pxf, Becker:2021ivq, Becker:2022wlo, Cardoso:2021wlq, PhysRevLett.129.241103, Cardoso:2022whc, PhysRevD.106.044027, Cole:2022yzw, CanevaSantoro:2023aol, Rahman:2023sof, Kadota:2023wlm, Zhang:2024hrq, Yue:2024xhf, Bertone:2024rxe, Zhao:2024bpp, Zwick:2024yzh, Zhou:2024vhk, Karydas:2024fcn,Cheng:2024mgl,tahelyani2024probingdarkmatterhalo}.

Similar to the single-electron wave function confined by a proton, ultralight bosonic field can be bound to a pointlike compact object, forming purely classical gravitational atom. The gravitational atoms (GAs) of spin 0,1,2 bosons can all naturally arise from the rotational superradiant (SR) instabilities of fast-spinning astrophysical BHs \cite{PhysRevD.22.2323,Dolan:2007mj,Arvanitaki:2010sy,Brito:2014wla,book_superradiance,Brito:2014wla,Baryakhtar:2017ngi,Frolov:2018ezx,Dolan:2018dqv,Siemonsen:2019ebd,Percival:2020skc,Siemonsen:2022ivj,PhysRevD.108.064001}. If the boson is sufficiently light, the hydrogenic description of the cloud will be a good approximation, since most part of the cloud is nonrelativistic (NR) and its wave dynamics is dominated by the gravity of the central BH. Being a NR structure, the hydrogenic GA may also be populated by the NR processes \cite{Budker:2023sex}.

In addition to the signatures of isolated GAs, the phenomenology of GA forming a binary with another companion body, referred as a \textit{GA-companion system}, has also attracted wide attention. The gravitational dynamics of GA-companion system was studied from several different aspects. In Refs. \cite{Ferreira:2017pth,Hannuksela:2018izj}, the motion of a test body in the gravitational background of scalar $|211\rangle$ cloud (the cloud of scalar GA in $|nlm\rangle=|211\rangle$ state, using the hydrogenic convention for the principal quantum number $n$) was investigated. Indeed, if the cloud is only weakly perturbed, the most important effect on the companion would be that due to the metric perturbations sourced by the whole cloud. In the Newtonian limit and within a sufficiently short timescale, it suffices to consider the stationary Newtonian potential of the cloud, which was adopted in Refs.~\cite{GRAVITY:2019tuf,Yuan:2022nmu,GRAVITY:2023cjt} and \cite{GRAVITY:2023azi} to derive constraints on the mass of scalar $|211\rangle$ cloud and vector $|1011\rangle$ cloud around Sagittarius A*. The presence of a companion can lead to tidal distortion of the cloud hence perturbing its Newtonian potential \cite{DeLuca:2021ite,Arana:2024kaz}. Even if the tidal distortion is negligible, this potential can be nonstationary due to the intrinsic processes in the cloud. In the case of real bosonic field, the cloud keeps annihilating into gravitational waves, leading to a secular depletion of the cloud mass. The resulting temporal evolution of gravitational potential can dominate the orbital evolution at sufficiently large radius, pushing the companion toward outspiral \cite{Kavic:2019cgk,Cao:2023fyv}. Atomic dynamics comes into play when the gravitational level mixings of GA become non-negligible. The oscillating tidal potential acting on GA by its companion leads to the possibilities of resonant transitions between the bound states \cite{B1, Zhang:2018kib, Berti_1,B3, Takahashi:2021yhy, Takahashi:2023flk, Tomaselli:2024bdd, Boskovic:2024fga, Tomaselli:2024dbw} (henceforth referred as the ``resonances'') as well as possible nonresonant effects \cite{WY_1,WY_2,WY_3,Fan:2023jjj}. Taking into account also the free states of GA, the induced bound-free transitions result in the cloud ``ionization'' \cite{B4,B5,Tomaselli:2023ysb}, whose backreaction provides a proper description for the dynamical friction (DF) \cite{ZJ,B4,B5,Tomaselli:2023ysb} experienced by the companion in the nonrelativistic regime. Going beyond the Newtonian approximation, fully relativistic investigation of circular extreme mass-ratio inspiral into a Schwarzschild BH surrounded by complex scalar cloud has been carried out using black hole perturbation theory in Refs.~\cite{Brito:2023pyl,Duque:2023cac}.

So far, most of the studies on GA-companion system have focused on the scalar GA. However, the cases of bosonic fields with nonzero spin are also of theoretical interest and thus warrant proper investigation. An ultralight vector field can exist as dark photon \cite{Holdom:1985ag} or have string theory origins \cite{Goodsell:2009xc}. Besides the production via SR, it can also arise from nonthermal processes in the early Universe \cite{Graham:2015rva, Ema:2019yrd, Ahmed:2020fhc, Long:2019lwl, PhysRevD.99.035036, Nakayama:2019rhg} and is thus a viable DM candidate. In such scenarios, the formation of GAs would provide a complementary signature independent of the large-scale observables. In our previous study \cite{Cao:2023fyv}, we estimated and compared the effects of DF and cloud depletion due to its intrinsic GW radiation on the orbital evolution of circular GA-companion systems for various spin 0,1,2 GA states. We found the cloud depletion-dominated outspiral phase is roughly unaffected by the DF and both effects may lead to detectable observational signatures in the GA-pulsar binaries. However, the DF model used in Ref.~\cite{Cao:2023fyv} is a qualitative one and thus inadequate for a concrete study of the inspiral process, during which DF may be significant.

In this paper, we model the off-resonant orbital dissipation more consistently, replacing the simple estimation of DF with ionization fluxes \cite{B4,Tomaselli:2023ysb}. Furthermore we take into account the conservative effect of cloud gravity, which generally contributes to the secular variations of angular osculating elements. We also correct the treatment of cloud depletion effect in Appendix~\ref{appendix_DC}. For the sake of concreteness as well as simplicity, we consider the vector SR ground state $|1011\rangle$, which has also been the main focus concerning the searches of vector GA \cite{Chen:2022kzv,PhysRevD.108.064001, Fell:2023mtf, GRAVITY:2023azi}. The adiabatic orbital evolution of small companions in the $|1011\rangle$ cloud is studied based on the perturbative model. We find that low-frequency eccentric IMRI systems can be good targets for the future space-based gravitational-wave detectors to probe the existence of such vector clouds.

This paper is organized as follows. After reviewing the basics of vector GA in Sec.~\ref{sec_2}, we present our perturbative model of GA-companion binary system in Sec.~\ref{sec:GAcompanion}, with a focus on the $|1011\rangle$ state. Then in Sec.~\ref{sec_3} we investigate the orbital evolution of IMRIs around a vector GA saturated in the $|1011\rangle$ state and the associated gravitational waveforms. We summarize and discuss the results in Sec.~\ref{sec_4}. Throughout our discussion, we use the natural units $\hbar=G=c=1$ and flat spacetime metric $\eta_{ab}=\text{diag}(-1,1,1,1)$.

\section{Vector GA}\label{sec_2}
We begin with a brief recapitulation of the basic properties of an isolated GA in the Newtonian limit. The atomic description is similar for the free massive bosonic fields with spin $s=0,1,2$ \cite{PhysRevD.22.2323,Baryakhtar:2017ngi,Brito_2020}, here we focus on the vector case. A vector field $A_a=(A_0,\mathbf{A})$ with particle mass $\mu$ is described by the Lagrangian
\begin{equation}
	\mathcal{L}=-\frac{1}{4}F_{ab}F^{ab}-\frac{1}{2}\mu^2A^aA_a,\quad F_{ab}=\partial_a A_b-\partial_b A_a.
\end{equation}
Taking the NR limit of the ansatz
\begin{equation}\label{wave function}
A_i=\frac{1}{\sqrt{2\mu}}\left(\Psi_i\,e^{-i\mu t}+\text{c.c.}\right)\sqrt{\frac{M_\text{c}}{\mu}},
\end{equation}
and considering a flat spacetime background with scalar metric perturbation: $ds^2=-(1+2\Phi)dt^2+(1-2\Phi)|d\mathbf{x}|^2$, the wave equation of $A_i$ reduces to the Shr\" {o}dinger equation
\begin{equation}\label{Schrodinger}
	i\partial_t\boldsymbol{\Psi}=-\frac{1}{2\mu}\nabla^2\boldsymbol{\Psi}+\mu\Phi\boldsymbol{\Psi}.
\end{equation}
In this approximation, the wave function in a static central potential $\Phi(r)$ possesses the conserved charges $\{M_\text{c},\, E_\text{c},\, \mathbf{S}_\text{c}\}
=\int d^3r\,
\{\rho,\, \epsilon, \, \mathbf{j}\}$, with the NR mass density $\rho$, the NR energy density $\epsilon$ and the NR angular momentum density $\mathbf{j}$ given by
\begin{align}
\rho &= M_\text{c}\,\boldsymbol{\Psi}\cdot\boldsymbol{\Psi}^*,
\\
\epsilon &= \left(\frac{1}{2\mu}\nabla\Psi_i\cdot\nabla \Psi_i^*+\mu\Phi\boldsymbol{\Psi}\cdot\boldsymbol{\Psi}^*\right)\frac{M_\text{c}}{\mu},
\label{epsilon}
\\
\mathbf{j} &= i\left(
\Psi_i^*\nabla \Psi_i\times\mathbf{r}
+\boldsymbol{\Psi}\times\boldsymbol{\Psi}^*
\right)\frac{M_\text{c}}{\mu}. \label{j}
\end{align}
The two terms in Eq.~\eqref{j} correspond to the orbital and spin angular momentum \cite{Jain:2021pnk}, respectively. Here we choose the normalization $\int d^3r\,\boldsymbol{\Psi}\cdot\boldsymbol{\Psi}^*=1$.

The bound state around a pointlike central body with mass $M$ and Newtonian potential $\Phi=-M/r$ has a hydrogenic structure. Choosing a reference $z$-direction, the basis of bound state can be labeled by four quantum numbers $|nljm\rangle$ as
\begin{equation}
	\boldsymbol{\Psi}_{nljm}(t,\mathbf{r})=R_{nl}(r)\,\mathbf{Y}_{ljm}(\theta,\phi)\,e^{-i\left[\omega^{(nljm)}-\mu\right]t},
	\label{hydrogen}
\end{equation}
where $R_{nl}(r)$ is the hydrogenic radial function and $\mathbf{Y}_{ljm}(\theta,\phi)$ the purely orbital vector spherical harmonics, with $n\ge 1$, $l\in[0,n-1]$, $j\in[|l-1|,l+1]$, $m\in[-j,j]$. For the scalar GA eigenstate $|nlm\rangle$, the angular function is replaced by the scalar spherical harmonics $Y_{lm}(\theta,\phi)$. The size of GA is measured by its Bohr radius $r_\text{c}\equiv M/\alpha^2$, with $\alpha\equiv \mu M$ being the gravitational fine structure constant. For convenience, we introduce the nondimensional radius $x\equiv r/r_\text{c}$ and choose the normalization $\langle nljm|n'l'j'm'\rangle=\int d^3r\,\boldsymbol{\Psi}_{nljm}\cdot\boldsymbol{\Psi}^*_{n'l'j'm'}=\delta_{nn'}\delta_{ll'}\delta_{jj'}\delta_{mm'}$, a nondimensional radial function can then be defined as $R_{nl}(x)\equiv r_\text{c}^{3/2}R_{nl}(r)$. The cloud occupied by a single eigenstate $|nljm\rangle$ has energy $E_\text{c}=\frac{M_\text{c}}{\mu}\left[\omega^{(nljm)}-\mu\right]$ and angular momentum $\mathbf{S}_\text{c}=\frac{M_\text{c}}{\mu}m \,\mathbf{e}_z$, where we define $M_\text{c}$ to be the total mass of the cloud. In the presence of multiple eigenstates, the wave function can be written as a linear superposition:\footnote{In the presence of perturbation, this decomposition of bound state becomes instantaneous, with $c_{nljm}$ being time-dependent.} $|\boldsymbol{\Psi}\rangle=\sum_{n,l,j,m} c_{nljm}|nljm\rangle$, with $E_\text{c}=\sum_{n,l,j,m} \frac{M_\text{c}}{\mu}\left[\omega^{(nljm)}-\mu\right]|c_{nljm}|^2$ and $\mathbf{S}_\text{c}\cdot\mathbf{e}_z=\sum_{n,l,j,m} \frac{M_\text{c}}{\mu}m|c_{nljm}|^2$, but $\mathbf{S}_\text{c}$ is not necessarily parallel to the $z$-direction, e.g., when the $|nljm\rangle$ and $|nlj'm'\rangle$ states coexist. Besides the bound state which constitutes the cloud, there are also free state spanned by
\begin{equation}
\boldsymbol{\Psi}_{k,ljm}(t,\mathbf{r})=R_{k,l}(r)\,\mathbf{Y}_{ljm}(\theta,\phi)\,e^{-i\left[\omega^{(k,ljm)}-\mu\right]t},
\end{equation}
(see \cite{B4} for the explicit form of $R_{k,l}$) with $\omega^{(k,ljm)}-\mu=-\frac{k^2}{2\mu}$ and $\lim_{r\to \infty}R_{k,l}(r)=\frac{2}{r}\sin(kr+\Upsilon)$. It represents the part of nonrelativistic bosonic field unbound to the central body. We choose the normalization $\langle k',l'j'm'|k,ljm\rangle=2\pi\,\delta(k-k')$ and introduce the nondimensional wave number $\tilde k=r_\text{c}k$, a nondimensional radial function can then be defined as $R_{\tilde k,l}(x)\equiv r_\text{c}R_{k,l}(r)$.

Thus far we have considered an exact hydrogenic solution, for which the energy spectrum in Eq.~\eqref{hydrogen} is given by $\omega^{(nljm)}=\mu-\frac{\mu\alpha^2}{2n^2}$. But even at the purely nonrealtivistic level, the cloud has a minimal self-interaction due to its own gravity, there can also be effects stemmed from the special relativistic correction and the possibly strong gravitational field near the central body. For a \textit{light} cloud with $\beta<\alpha$ and in the \textit{Newtonian limit}: $\alpha\ll 1$, since $r\sim r_\text{c}\gg M$, these would manifest mainly as a small correction $\Delta \omega^{(nljm)}=\Delta \omega^{(nljm)}_\text{rel}+\Delta \omega^{(nljm)}_\text{self}$ to the energy level\footnote{The metric perturbations sourced by the cloud is discussed in Appendix~\ref{appendix_metric perturbations}. For the ionization process to be considered later, only the Bohr structure of the energy spectrum is relevant, hence these corrections can be safely neglected in the Newtonian limit.}. The relativistic correction \cite{B2} $\Delta \omega^{(nljm)}_\text{rel}\sim \mu \alpha^4$, the self-gravity correction from the gravitational potential of the cloud $\Delta \omega_{\text{self},\Phi}^{(nljm)}\sim \mu \alpha^2\beta$, and the correction from gravitomagnetic field of the cloud $\Delta \omega_{\text{self},\Xi}^{(nljm)}\sim\mu \alpha^4\beta$, where $\beta\equiv M_\text{c}/M$ is the mass ratio between the cloud and the central body. In this paper we focus on the GA with SR origin for which the central body is a spinning BH, and $\Delta\omega^{(nljm)}_\text{rel}$ (with $z$-axis parallel to the BH spin) is small but crucial. $\Delta\omega^{(nljm)}_\text{rel}$ is generally complex (such that $\omega = \omega_R + i \omega_I$), when it acquires a positive imaginary part, the boson cloud grows spontaneously. Such an instability can be naturally realized near the spinning BH for a moderate value of $\alpha$. Once formed, the boson cloud is expected to be dominated by the mode with the fastest growth rate, for vector GA this is the $|1011\rangle$ state \cite{Baryakhtar:2017ngi,B3}. The growth saturates when the SR instability vanishes (i.e., $\omega_I=0$, the explicit condition is $\omega_R=m\Omega_H$, with $\Omega_H=\frac{1}{2M}\frac{\chi}{1+\sqrt{1-\chi^2}}$), and the vector cloud keeps depleting into gravitational waves. We consider only the regime $\alpha\ll 1$, in which the picture of hydrogenic GA is well justified. A large value of $\alpha$ necessitates relativistic treatment \cite{Fell:2023mtf,Cannizzaro:2023jle,Brito:2023pyl,Duque:2023cac,Speeney:2024mas} and is beyond the scope of the present work.

\section{Perturbative model of GA-companion system}\label{sec:GAcompanion}
If the GA and a companion object form a binary, there are inevitable gravitational interactions between the companion and the cloud. In the Newtonian limit, the nonrelativistic energy, linear and angular momentum will be exchanged between the bosonic field and the point masses, while the NR mass and spin angular momentum of the bosonic field are conserved, since the equation of motion of all components of $\boldsymbol{\Psi}$ are same, as given by Eq.~\eqref{Schrodinger}.

\begin{figure}[htb!]
	\centering
	\includegraphics[width=0.48\textwidth]{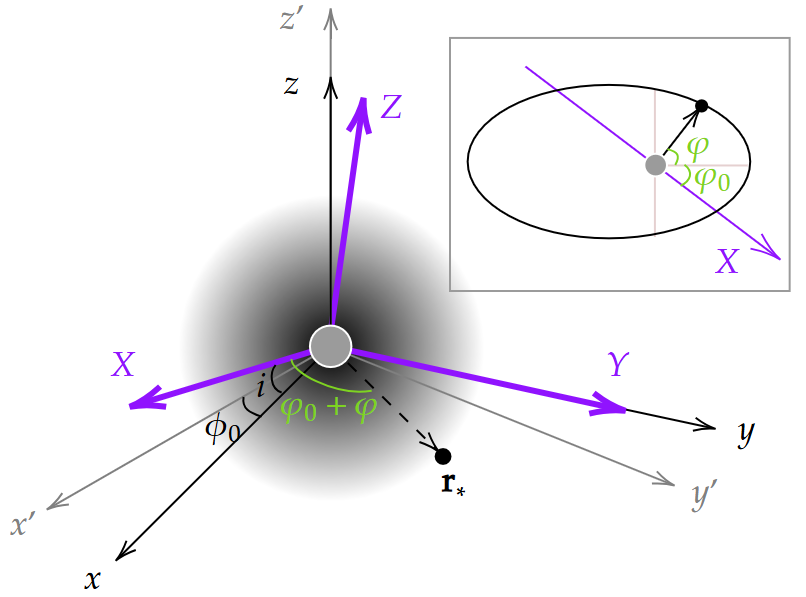}
	\caption{Definition of coordinate frames and the binary's angular orbital elements. In the $x'y'z'$ frame with the $z'$ axis parallel to the BH spin, the spherical coordinate of the companion is $(r_*,\theta_*,\phi_*)$, and we denote its coordinate in the $XYZ$ frame with $Z$ axis parallel to the orbit normal to be $(r_*,i,\varphi_0+\varphi)$, where $i$ is the inclination angle of the orbital plane relative to the BH equatorial plane, $\varphi$ the true anomaly, and $\varphi_0$ the argument of periastron on the orbital plane. The line of ascending node lies on the negative $y=Y$ axis. The $xyz$ frame is formed by rotating the $x'y'z'$ frame counterclockwise by $\phi_0$ about the $z'=z$ axis.}\label{orbit_2}
\end{figure}

We denote the position of GA's host BH with mass $M$ by $\mathbf{R}_1$ and the position of its companion with mass $M_*$ by $\mathbf{R}_2$. The two-body separation is $\mathbf{r}_*\equiv \mathbf{R}_2-\mathbf{R}_1$ with $r_*\equiv|\mathbf{r}_*|$, and the total mass is $M_\text{tot}\equiv M+M_*=(1+q)M$, with the mass ratio $q\equiv M_*/M$. Since we are mainly interested in the case of small mass ratio, the approximation $q\ll 1$ will be made in the following unless stated otherwise. To track the evolution of osculating orbit, we set up three BH-centered Cartesian coordinate frames as depicted in Fig.~\ref{orbit_2}, keeping the same convention with Ref.~\cite{Cao:2023fyv}.

The Keplerian elliptical orbit of a bound Newtonian binary in vacuum can be parametrized by six orbital elements 
$\{a,e,\varphi_0,\phi_0,i,t_0\}$ as defined in Fig.~\ref{orbit_2}, with
\begin{align}
r_*&=\frac{p}{1+e\cos\varphi}=a(1-e\cos \xi),
\\
\mathcal{M}&=\xi-e\sin \xi=\Omega (t-t_0).
\end{align}
Here $a$ is the semimajor axis, $e\in[0,1)$ the orbital eccentricity, $p=a(1-e^2)$ the semilatus rectum, $\varphi$ the true anomaly, $\mathcal{M}$ the mean anomaly, $\xi$ the eccentric anomaly, $t_0$ the periastron-crossing time and $\Omega=\sqrt{M_\text{tot}/a^3}$ the mean orbital frequency. The orientation of the periastron on the orbital plane is measured by $\varphi_0$ and the orientation of the orbital plane is measure by $\phi_0$.

Under a perturbing acceleration $\mathbf{F}=F_r\mathbf{e}_r+F_\varphi \mathbf{e}_\varphi+F_Z\mathbf{e}_Z$, where $\mathbf{e}_\varphi=\mathbf{e}_Z\times \mathbf{e}_r$, the orbit evolves according to $\ddot{\mathbf{r}}_*=-(M_\text{tot}/r_*^2)\,\mathbf{e}_r+\mathbf{F}$, the resultant variation of osculating orbital elements is given by the Gaussian perturbation equation \cite{klioner2016basiccelestialmechanics,Blas:2021mpc,will2018theory}
\begin{align}
	\frac{\dot{a}}{a}&=\frac{2}{\Omega}\left(\frac{e\sin\varphi}{a\sqrt{1-e^{2}}}F_{r}+\frac{\sqrt{1-e^{2}}}{r_*}F_{\varphi}\right),
	\\
	\dot{e}&=\frac{\sqrt{1-e^2}}{a\Omega} \left[(\cos\varphi+\cos\xi)F_\varphi+\sin\varphi\, F_r\right],
	\\
	\dot{\varphi}_{0}&=\frac{\sqrt{1-e^2}}{ae\Omega}\left[\left(1+\frac{r_*}{p}\right)\sin\varphi F_\varphi-\cos\varphi\, F_r\right]-\cos i\,\dot\phi_0,
    \label{periastron precession}
	\\
	\dot \phi_0&=\frac{r_*\cos (\varphi+\varphi_0) F_Z}{a^2\Omega\sqrt{1-e^2}\sin i}=\frac{\Omega \left(a/r\right)^2\sqrt{1-e^2}-\dot\varphi-\dot\varphi_0}{\cos i},\label{orbit plane precession}
	\\
	\dot i&=-\frac{r_*\sin (\varphi+\varphi_0) F_Z}{a^2\Omega\sqrt{1-e^2}}, \label{i changing rate}
	\\
	\dot{\mathcal{M}} &=
	\Omega+\frac{a \cos \xi\,\dot e-(1-e\cos \xi)\dot a}{ae \sin\xi\,/\,(1-e\cos \xi)}-\sin\xi \,\dot e
	.
\end{align}
The Gaussian perturbation equation is exact, for the secular change of an orbital element $\mathcal{X}$ we consider only its first-order approximation given by the time average of $\dot{\mathcal{X}}$ within one orbit period $T=2\pi/\Omega$: $\langle \dot{\mathcal{X}} \rangle\equiv \frac{1}{T}\int_0^T dt\,\dot{\mathcal{X}}$, with the orbital elements in the integrand fixed to constant values.

For a Newtonian elliptical binary without dissipation, the conserved orbital energy and angular momentum are given by\footnote{The Laplace-Runge-Lenz vector $\mathbf{A}=\mathbf{v}\times (\mathbf{L}/\mu')/M_\text{tot}  -\mathbf{e}_r$ is also conserved. In the presence of $\mathbf{F}$, $\dot{\mathbf{A}}=[\mathbf{ F} \times (\mathbf{L}/\mu')+\mathbf{v} \times (\mathbf{r}\times \mathbf{ F}]/M_\text{tot}$, from this we can obtain the periastron precession rate given by Eq.~\eqref{periastron precession}.}
\begin{equation}
\begin{aligned}
E &=\mu'\left(\frac{v^2}{2}-\frac{M_\text{tot}}{r}\right)=-\frac{qM^2}{2a},
\\
\mathbf{L} &=\mu' \mathbf{r}\times\mathbf{v}=L\,\mathbf{e}_Z,\quad L=\frac{qM}{\sqrt{1+q}}\sqrt{(1-e^2)aM},
\end{aligned}
\end{equation}
where $\mu'=MM_*/M_\text{tot}$ is the reduced mass. In the presence of $\mathbf{F}$, the evolution of $E$ and $\mathbf{L}$ of the osculating orbit is given by $\dot E=\mathbf{v}\cdot (\mu'\mathbf{F})$ and $\dot{\mathbf{L}}=\mathbf{r}\times(\mu'\mathbf{F})$. The adiabatic evolution of $a$ and $e$ is related to the time-averaged energy flux $P\equiv -\langle\dot E\rangle$ and angular momentum flux normal to the orbital plane $\tau_Z\equiv\boldsymbol{\tau}\cdot\mathbf{e}_Z\equiv-\langle\dot L\rangle$ (with $\boldsymbol{\tau}\equiv -\langle \dot{\mathbf{L}}\rangle$) through
\begin{align}
\langle\dot e\rangle=\frac{1-e^2}{2e}\left(\frac{P}{E}+\frac{2\tau_Z}{L}\right),
\quad
\langle\dot a\rangle=\frac{aP}{E}.
\end{align}

We now consider the effects of the boson cloud on the binary's orbital evolution, assuming the cloud is dominated by a single bound eigenstate and remains nearly unperturbed by the presence of the small companion. At the Newtonian order, the orbital dynamics is given by $\mathbf{F}=-\nabla \Phi$, where $\Phi(t,\mathbf{r})$ is the gravitational potential sourced by the bosonic field and $\nabla\Phi$ is evaluated at $\mathbf{R}_2$. Meanwhile, the cloud is also influenced by the companion's gravity. Perturbatively, the dynamics of this system can be separated into the following two aspects: (i) the GA transitions induced by the companion and its backreaction due to the gravitational potential of the perturbed part of the cloud, (ii) the perturbing acceleration acting on the companion by the gravitational potential of the unperturbed cloud. These will be discussed in Sec.~\ref{(i)} and \ref{(ii)}, respectively. For convenience, we introduce the nondimensional quantities $x_a\equiv a/r_\text{c}$ and $x_*\equiv r_*/r_\text{c}$.

\subsection{Dissipative effects}\label{(i)}
Neglecting the contribution of the host BH's orbital motion to the energy and angular momentum of the cloud, the possible absorption of the cloud by the companion, and the perturbative effect of cloud gravity on the binary's conservative dynamics (to be discussed in the next section), the backreaction from the tidally induced atomic transitions can be read from the energy/angular momentum balance between the cloud and binary orbit:
\begin{equation}\label{balance}
\left\langle(\dot{\mathbf{S}}_\text{c}+\dot{\mathbf{J}})+\dot{\mathbf{L}}\right\rangle=-\boldsymbol{\tau},
\quad
\left\langle(\dot E_\text{c}+\dot M_\text{c}+\dot M)+\dot E\right\rangle=-P,
\end{equation}
where $\mathbf{J}=J\,\mathbf{e}_z$ is the spin angular momentum of the host BH and $E_\text{c}$ ($\mathbf{S}_\text{c}$) are the cloud's energy (angular momentum) evaluated in the BH-centered frame. The energy and angular momentum of the cloud and its host BH can be exchanged through the SR process, but this has no influence on the orbital dynamics if we neglect the effects from the resulting change of cloud gravity. The leading-order energy and angular momentum fluxes due to the binary's GW radiation are \cite{PhysRev.131.435}
\begin{align}
P_\text{gw}&=\frac{32}{5}\frac{(1+q)\,q^2\alpha^{10}}{x_a^5}f(e)
,\label{P_gw}
\\
\tau_\text{gw}&=\frac{32}{5}\frac{(1+q)^{1/2}q^2\alpha^7}{x_a^{7/2}}M\,h(e)
,\label{tau_gw}
\end{align}
with $f(e)=\frac{1+{73}e^2/{24}+{37}e^4/{96}}{(1-e^2)^{7/2}}$ and $h(e)=\frac{1+{7}e^2/{8}}{(1-e^2)^2}$.
Meanwhile, radiation of the bosonic field is excited by the oscillating tidal potential of the companion, leading to the ionization of GA. Similar to the GW damping, the secular backreaction of the ionization process can be more easily described with the time-averaged energy flux $P_\text{ion}$, angular momentum flux $\boldsymbol{\tau}_\text{ion}=\tau_{\text{ion},X}\,\mathbf{e}_X+\tau_{\text{ion},Y}\,\mathbf{e}_Y+\tau_{\text{ion},Z}\,\mathbf{e}_Z$ and mass flux $(\dot M_\text{c})_\text{ion}$. Details on the computation of these fluxes are explained in Appendices~\ref{appendix_mixings} and \ref{appendix_ionization}.

In Eq.~\eqref{balance}, we can neglect the component of $\mathbf{S}_\text{c}$ perpendicular to the $z$-axis if it stays conserved throughout the possible transitions. In the case of scalar GA, since the angular momentum operators $\hat L_\pm\equiv \hat L_x\pm i\hat L_y$ (with $\hat{\mathbf{L}} \equiv -i\mathbf{r} \times \nabla$) satisfy $\hat L_\pm |lm\rangle=\sqrt{l(l+1)-m(m\pm 1)}\,|l,m\pm 1\rangle$, $\mathbf{S}_\text{c}$ can deviate from the $z$-direction when adjacent hyperfine levels coexist\footnote{This may arise, e.g., in a scalar $|211\rangle$ cloud due to the hyperfine transition $|211\rangle \leftrightarrow |210\rangle$ through $l_*=1$ and (for inclined orbit) $l_*=2$.}. The situation is similar for the vector GA, since only the orbital part in \eqref{j} is responsible for the backreaction. For $\mathbf{S}_\text{c}=S_\text{c}\,\mathbf{e}_z$, the orbital evolution is given by
\begin{align}
		P_\text{eff}&=P_\text{gw}+P_\text{ion}+\langle\dot E_\text{c}+\dot M_\text{c}+\dot M\rangle \label{dot_E}
		,
		\\
		\tau_\text{eff}&=\tau_\text{gw}+\tau_{\text{ion},Z}+\langle\dot S_\text{c}+\dot J\rangle \cos i, \label{dot_L}
		\\
		\langle\dot i\rangle&=\frac{1}{L}\left[\tau_{\text{ion},X}+\langle\dot S_\text{c}+\dot J\rangle\sin i\right], \label{dissip_i}
        \\
        \langle\dot \phi_0\rangle&=\frac{1}{L \sin i}\,\tau_{\text{ion},Y}, \label{dissip_phi0}
		\\
		\langle\dot e\rangle&=-\left(\frac{1-e^2}{e}\right)\frac{aP_\text{eff}}{qM^2}\left(1-\frac{\tau_\text{eff}\,\Omega}{P_\text{eff}\sqrt{1-e^2}}\right), \label{dissip_e}
		\\
		\langle\dot x_a\rangle&=-\frac{2}{qM\alpha^2}\,x_a^2\,P_\text{eff},
        \\
        \langle \dot M_\text{c} \rangle&= -P_\text{gw,c}+(\dot M_\text{c})_\text{ion},
        \label{dissip_a}
\end{align}
where $\dot S_\text{c}+\dot J$, $\dot E_\text{c}+\dot M_\text{c}+\dot M$ is associated with the change of cloud due to the possible bound state transitions induced by the companion\footnote{Note this corresponds to the evolution of $c_i$ in the expansion $|\psi(t)\rangle=\sum_i c_i |i\rangle$, rather than the evolution of $c_i|i\rangle$.}, while the contributions from bound-free transitions to $\dot S_\text{c}$ and $\dot E_\text{c}$ have been absorbed into the ionization fluxes. $P_\text{gw,c}$ is the GW energy flux of the cloud. Being dissipative effects, the ionization-induced inclination change and orbital precession are typically negligible compared with the conservative effects of the cloud. In the case of a spherical ($l=0$) state being ionized, we simply have $\tau_{\text{ion},X}=\tau_{\text{ion},Y}=0$.

For the ionization of the $|1011\rangle$ cloud, instead of computing directly with the vector wave functions, it is much more convenient to perform a scalar field reduction. The ionization flux of a vector GA in the $|1011\rangle$ state is same with that of a scalar GA in the hydrogenic ground state $|100\rangle$, since their wave functions are simply related by
\begin{equation}
\boldsymbol{\Psi}_{1011}=-\frac{1}{\sqrt{2}}\left(\begin{matrix}
1\\ i \\ 0
\end{matrix}\right)\Psi_{100},
\end{equation}
where $\Psi_{100}(t,\mathbf{r})=r_\text{c}^{-3/2}\frac{e^{-x}}{\sqrt{\pi}}\,e^{i\mu\frac{\alpha^2}{2}t}$. This equivalence can also be straightforwardly confirmed by the numerical computation. Moreover, since the $|100\rangle$ state is spherically symmetric, the ionization fluxes are independent of $\{i,\varphi_0,\phi_0\}$. Without loss of generality, for the computation we can set $i=\varphi_0=\phi_0=0$, the selection rule of scalar GA from the radial integral in $\langle k,l'm'|V_*|nlm\rangle$ is \cite{B3} $m'=m+m_*\in[-l',l']$, $|m_*|\le l_*\in[|l'-l|,l'+l]$ with $l_*+l+l'$ being even. The ionization fluxes are contributed by the final states $|k,l'm'\rangle$ with nonvanishing transition amplitudes as given by this selection rule. In the present case, choosing $|nlm\rangle=|100\rangle$ leaves the only possibility $l'=l_*$ and $m'=m_*$, though we still need to sum over a large set of $(l',m')$ states in order to achieve good convergence in the case of high eccentricity. Note such a scalar field reduction cannot be applied to the bound-state mixing when the correction to Bohr structure becomes important, see Appendix~\ref{appendix_Bohr} for further discussions on the level mixings with $|1011\rangle$ state.

\begin{figure}[t]
	\centering
	\includegraphics[width=0.48\textwidth]{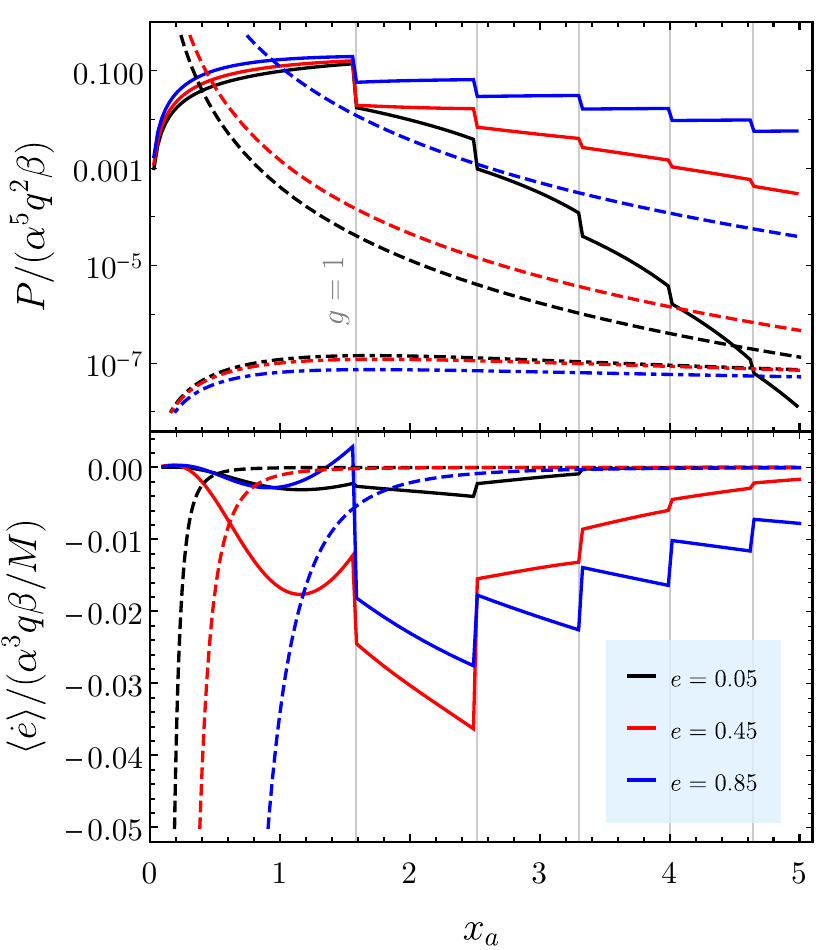}
	\caption{Dissipative energy flux and the rate of change of eccentricity due to ionization (solid lines) and 2.5PN GW damping (dashed lines), and the effective power $|P_\text{DC}|$ arising from the GW depletion of the cloud (dot-dashed line in the upper panel) for eccentric orbit in the vector $|1011\rangle$ cloud, with $\alpha=10\beta=0.05$ and $q=10^{-4}$. Note that $P_\text{gw}\propto \alpha^{10}q^2$ and $\langle\dot e\rangle_\text{gw}\propto \alpha^8 q/M$.}\label{P_eccentric}
\end{figure}

\begin{figure*}[htb]
	\centering
	\includegraphics[width=0.48\textwidth]{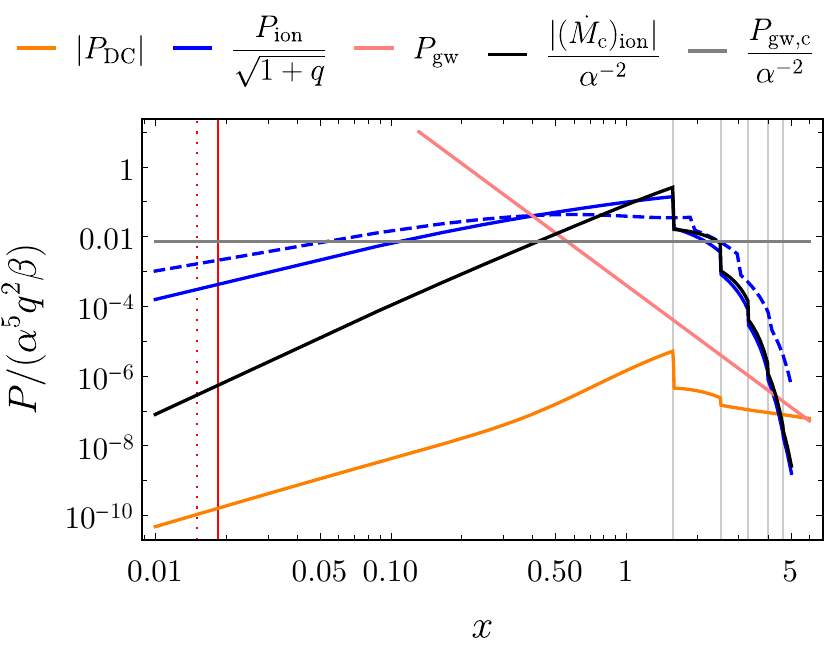}
	\quad
	\includegraphics[width=0.47\textwidth]{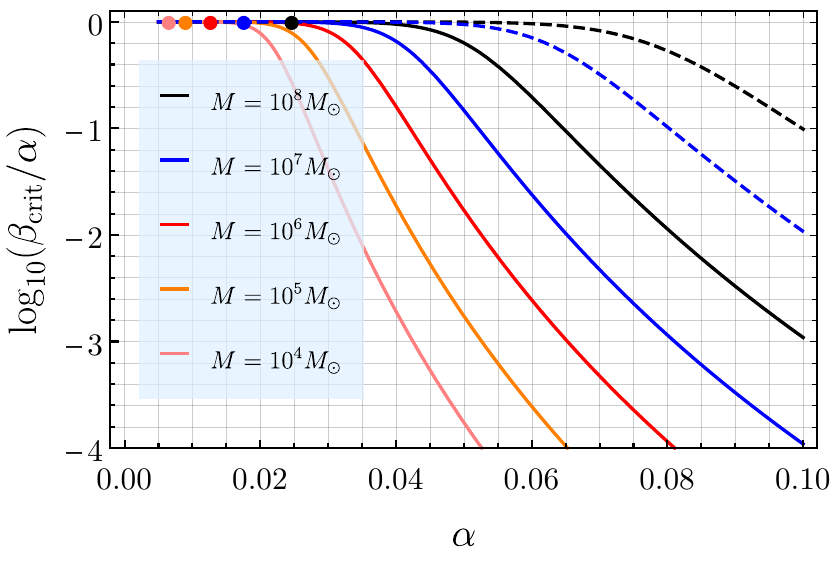}
	\caption{\textbf{Left:} comparison between various contributions to the effective dissipation power of orbital energy for circular orbit in the vector $|1011\rangle$ cloud, with $\alpha=10\beta=0.05$ and $q=10^{-4}$ (solid lines) or $q=0.8$ 
 (dashed line). The vertical red solid line marks $x=\alpha^{4/3}$ and the vertical red dotted line marks $x=6\alpha^2$, both are deep inside the cloud and beyond the regime of interest. \textbf{Right:} the mass of $|1011\rangle$ cloud with a given age for $\chi_i=1$. Solid (dashed) lines correspond to $\tau_\text{c}=10^8\,(10^6)$ yr. In comparison, the minimum value of $\alpha$ satisfying $\tau_I(\chi=1)\le 10^{6}\,\text{yr}$ for a given BH mass is indicated by the solid point, here $\tau_I=1/(2\omega_I)$ is the timescale of the initial exponential growth of the cloud mass during the SR process.}\label{1011_timescale}
\end{figure*}

One of the main features of the ionization flux as a function of $x_a$ is the presence of sharp jumps \cite{B4}. As explained in Appendix~\ref{appendix_ionization}, a bound state with energy level $\epsilon_n=-\mu\alpha^2/(2n^2)$ is ionized into the free states with wave number $\tilde k^{(g)}\equiv\sqrt{2g(1+q)^{1/2}x_a^{-3/2}-n^{-2}}$, where $g\in \mathbb{Z}^+$ and $\Omega=\mu\alpha^2(1+q)^{1/2}x_a^{-3/2}$ is the mean orbital frequency. The jump appears at the outset of an ionization process (corresponding to a final state with $k=0$) due to the finite bound-free level mixing at $k=0$. In reality, the jump should be smoothed by the transient process before the ionization approaches the final steady state \cite{B4}, only the latter is described by the ionization flux. This condition $k=0$ gives the semimajor axis of the jump: $x^{(g)}\equiv\left(2g\right)^{2/3}n^{4/3} (1+q)^{1/3}$, which are independent of the orbital eccentricity. For $q\ll 1$, $\tilde k^{(g)}$ depends solely on $x_a$, which leads to the simple scaling relations: $P_\text{ion}\propto q^2\alpha^5$, $(\dot M_\text{c})_\text{ion}\propto q^2\beta\alpha^3$, $\tau_{\text{ion},Z}\propto q^2\beta \alpha^2M$ and $\langle \dot e\rangle_\text{ion}\propto \alpha^3 q\beta/M$. Since they share the same $q$-scaling with the 2.5 post-Newtonian (PN) order effects given by Eqs.~\eqref{P_gw} and \eqref{tau_gw}, evolution of $\{x_a,e\}$ under the GW damping and ionization will be nearly independent of the binary mass ratio if $q\ll 1$. When $q$ is large, the ionization flux is boosted by the factor $q^2$, but the enhancement is actually more significant at large radius (see Fig.~\ref{1011_timescale}). In such a case the perturbative treatment would be valid only at very large distance from the cloud. For $q\sim 1$, the cloud can possibly form a delocalized structure akin to the electron wave function in a hydrogen molecular ion, but its ionization during the secular orbital evolution may also be significant.

\begin{figure}[htb]
	\centering
	\includegraphics[width=0.47\textwidth]{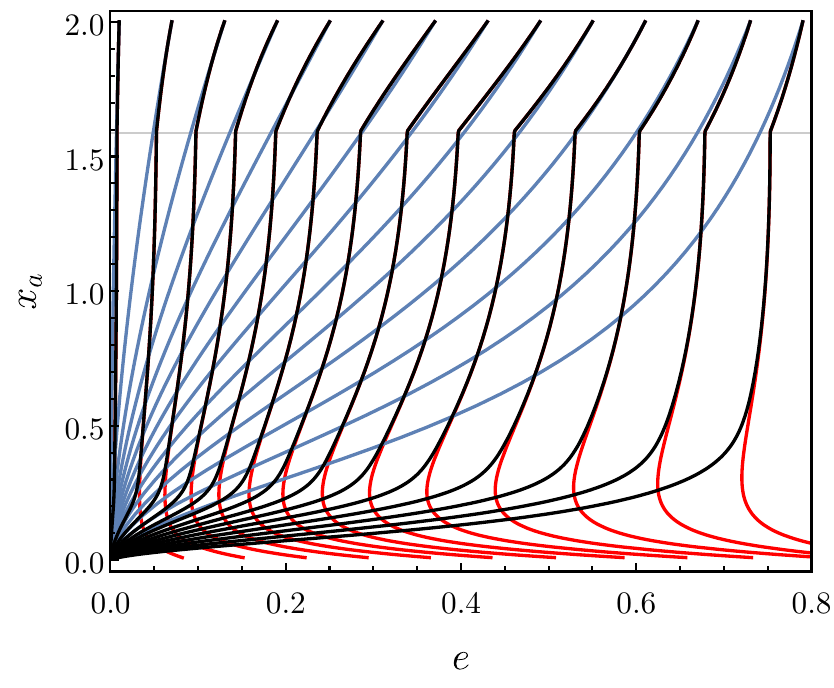}
	\caption{Evolution flow of $\{x_a,e\}$ in the vector $|1011\rangle$ cloud under $P_\text{gw}$ (blue lines), $P_\text{ion}$ (red lines) and $P_\text{gw}+P_\text{ion}$ (black lines), for $\alpha=0.03$, $\beta=0.01$ and $q=10^{-4}$. The trajectories start at $x_a=2$, with $x^{(1)}$ indicated by the gray horizontal line.}\label{flow}
\end{figure}

The dissipation power and the rate of change of eccentricity due to the ionization of vector $|1011\rangle$ cloud for three different orbital eccentricities in the small $q$ limit are shown in Fig.~\ref{P_eccentric}. For comparison, we also plot the results of GW damping and the cloud depletion effect (see Appendix~\ref{appendix_DC}). $P_\text{ion}$ is found to be a monotonically increasing function of $e$, and so is $\langle \dot e \rangle_\text{ion}$ for sufficiently large semimajor axis. The ionization power $P_\text{ion}$ peaks at the first ionization jump $x^{(1)}\approx 1.6$, and increase (decrease) with $x_a$ below (above) $x^{(1)}$ unless for a very large eccentricity, in which case $P_\text{ion}$ slightly increases between the higher-order consecutive jumps. $\langle \dot e\rangle_\text{ion}$ typically becomes positive at small enough radius $(x_a\sim 0.1)$ and also at an intermediate range of $x_a$ when the eccentricity is high. But in all cases we have examined, the orbit will not undergo eccentrification due to the combined effects of GW damping. The GW flux of the binary are suppressed relative to the ionization flux by a factor of $\alpha^5/\beta$ and can be subdominant during the early stage of inspiral, the minimum semimajor axis of this regime is shifted away from the cloud as the eccentricity increases, below which the orbital decay approach the vacuum trajectories, as depicted in Fig.~\ref{flow}. A notable feature of the ionization-dominated evolution is that, even if $|\langle\dot e\rangle|$ is increased relative to the vacuum case, the eccentricity decays more slowly with respect to $x_a$. The binary is thus more eccentric at a given orbital frequency for the same initial condition.

\subsection{Conservative effects}\label{(ii)}
In addition to the dissipative effects, the Keplerian motion also receives conservative perturbation from relativistic corrections and the cloud gravity. In the regime of interest, the primary ones are\footnote{Here we take $-\nabla\Phi|_{\mathbf{r}=\mathbf{R}_1}=\mathbf{0}$, hence the relative acceleration induced by $\Phi$ is $-\nabla\Phi|_{\mathbf{r}=\mathbf{R}_2}$. This is the case if the cloud is occupied by a single eigenstate in the comoving frame of its host BH (see Appendix~\ref{appendix_mixings}). Note that we are considering only the motion of the host BH and its companion. In the limit of large orbital radius, it will be a better approximation to treat the GA as a single body with mass $M+M_\text{c}$.}
\begin{equation}\label{conservative_F}
	\mathbf{F}\approx
\mathbf{F}_{\mathrm{1PN}}
-\nabla\Phi+\mathbf{v}\times[\nabla\times(\boldsymbol{\Xi}+\boldsymbol{\Xi}_\bullet)],
\end{equation}
where $\mathbf{F}_{\mathrm{1PN}}$ is the leading first post-Newtonian (1PN) correction to the binary dynamics \cite{Blanchet:2013haa}
\begin{equation}
\begin{aligned}
\frac{\mathbf{F}_{\mathrm{1PN}}}{F_\text{0PN}}=& \left[(4+2\nu)\,\frac{M_\text{tot}}{r_*}-(1+3\nu)\,v^{2}+\frac{3}{2}\nu\,{\dot{r}_*}^{2}\right]\mathbf{e}_r\\
&+(4-2\nu)\,\dot r_*\,\mathbf{v},
\end{aligned}
\end{equation}
with $F_\text{0PN}=M_\text{tot}/r_*^2$, $\mathbf{v}\equiv \dot{\mathbf{r}}_*\equiv v\,\mathbf{e}_\varphi$, and $\nu\equiv q/(1+q)^2$ the symmetric mass ratio. $\Phi$ is the stationary gravitational potential of the cloud,
\begin{equation}
\boldsymbol{\Xi}=-\frac{2J_\text{c}(\mathbf{r})\,\sin \theta}{r^2}\mathbf{e}_{\phi}
,\quad
\boldsymbol{\Xi}_\bullet=-\frac{2J\sin \theta}{r^2}\mathbf{e}_{\phi}
,
\end{equation}
are the stationary gravitomagnetic potentials sourced by the cloud and the spin $J=M^2\chi$ of the host BH, respectively. The explicit expressions of $\Phi$ and $\boldsymbol{\Xi}$ are derived in Appendix~\ref{appendix_metric perturbations}. For a cloud saturated in the state with azimuthal quantum number $m$, the BH spin is reduced to $\chi\approx 4\alpha/m$. For $\beta \gg \alpha^2$, the gravitomagnetic field of GA is then dominated by the cloud at sufficiently large radius, because it carries most part of the angular momentum but only a small fraction of the total mass.

\begin{figure}[htb]
	\centering
	\includegraphics[width=0.45\textwidth]{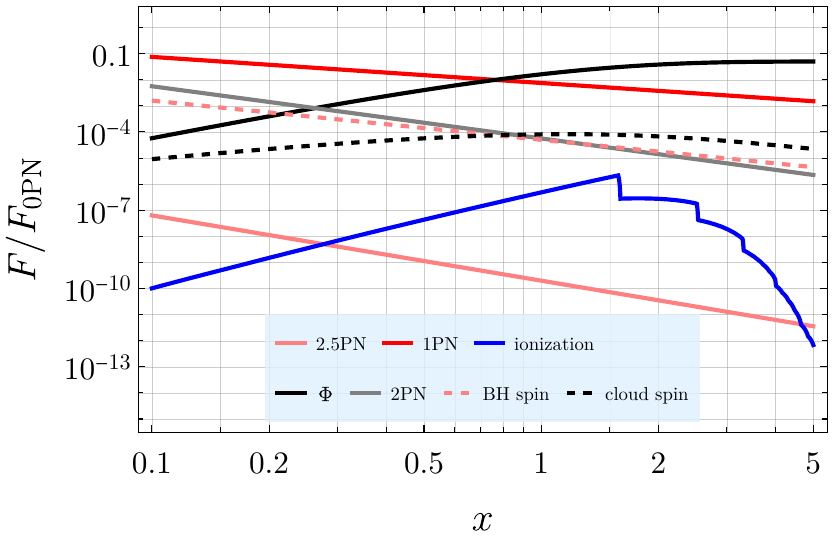}
	\caption{Magnitudes of the 1PN gravitoelectric (red solid line), 1.5PN gravitomagnetic (red dashed line), 2.5PN  dissipative (pink line) accelerations due to the central BH, and the accelerations due to the stationary cloud gravitational potential (black solid line), gravitomagnetic field (black dashed line) and ionization (blue line) relative to $F_\text{0PN}$ for equatorial circular orbit in the vector $|1011\rangle$ cloud, with $\alpha=\beta=0.05$ and $q=10^{-4}$. For the backreaction of ionization, we plot $F=P_\text{ion}/(M_*v)$.}\label{force_circular}
\end{figure}

\begin{figure}[htb]
	\centering
	\includegraphics[width=0.45\textwidth]{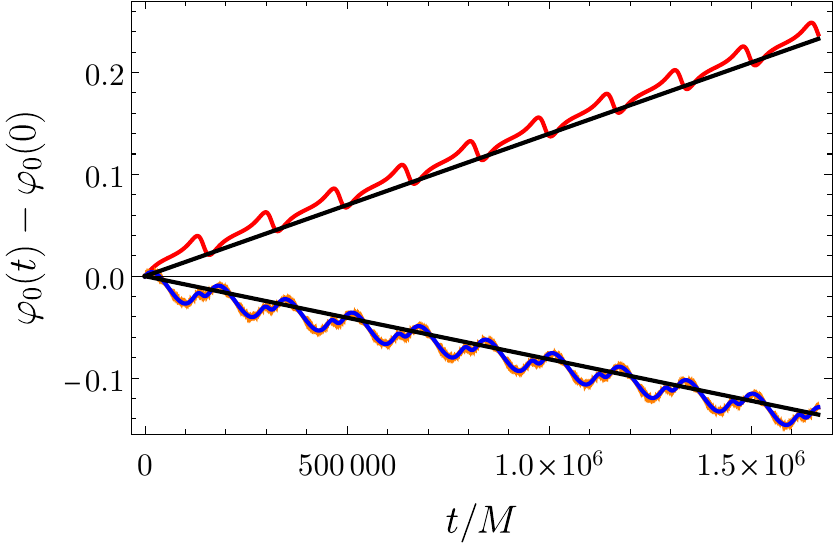}
	\caption{Comparison between the exact evolution of osculating element $\varphi_0$ (colored lines) and the analytical approximation (black lines) for its secular drift, for orbit with initial condition $x_a(0)=0.8$, $e(0)=0.3$, $\varphi_0=0.5$, $\phi_0=3$, $i=\pi/3$ and the parameters $\alpha=\beta=0.03$, $q=10^{-4}$. Blue (red) line corresponds to the result with (without) stationary gravitational potential of the cloud (in both cases $\mathbf{F}_\text{1PN}$ is included).  The result taking into account the oscillatory term in Eq.~\eqref{Phi_full} is depicted as the orange line, which displays only small and fast oscillations around the blue line.}\label{precession_check}
\end{figure}

In Fig.~\ref{force_circular}, the magnitudes of these conservative forces and some higher-order perturbations are compared with the dissipative effects for equatorial circular Keplerian orbits in the vector $|1011\rangle$ cloud. The backreaction of ionization in the case of a small mass ratio is seen to be much weaker than the stationary cloud gravity, justifying the perturbative treatment.

The conservative perturbation also contributes to the secular evolution of osculating orbit, as described by the Gaussian perturbation equation. $\mathbf{F}_\text{1PN}$ gives rise to the Schwarzschild precession
\begin{equation}
\langle\dot\varphi_0\rangle_\text{S}=\frac{3M_\text{tot}\Omega}{a(1-e^2)}=\frac{3(1+q)\alpha^5}{x_a^{5/2}(1-e^2)M}.
\end{equation}
while the gravitomagnetic acceleration due to the spin-orbit coupling with the central BH leads to the Lense-Thirring (LT) precession
\begin{equation}
\langle\dot\phi_0\rangle_\text{LT}=\langle\dot\phi_0\rangle_{\Xi_\bullet}=\frac{2\chi\alpha^6}{x_a^3(1-e^2)^{3/2}M}.
\end{equation}
This result corresponds to the limit $q\ll 1$. Under the same assumption it will be a good approximation to neglect the rotation of BH spin axis as well as the spin of the companion. The BH spin also contribiutes to the periastron precession, with $\langle \dot\varphi_0 \rangle_{\Xi_\bullet}=-3 \langle \dot\phi_0 \rangle_\text{LT}\,\cos i$.

The acceleration $-\nabla \Phi+\mathbf{v}\times(\nabla\times\boldsymbol{\Xi})$ (for time-independent $\Phi$ and $\boldsymbol{\Xi}$) is conservative, hence $\langle \dot a\rangle_\Phi=\langle \dot a\rangle_\Xi=0$. Additionally, the stationary Newtonian gravitational potential of GA (occupied by a single eigenstate) is symmetrical about the $z$-axis, thus $L_z=\mathbf{L}\cdot\mathbf{e}_z$ stays constant, from $L=L_z/\cos i\propto \sqrt{1-e^2}$ we obtain
\begin{equation}
\langle\dot e\rangle_\Phi/\langle\dot i\rangle_\Phi=(e-1/e)\tan i.
\end{equation}
The contribution of $\Phi$ and $\boldsymbol{\Xi}$ to the variation of angular osculating elements can be written as
\begin{align}
\langle \dot{\mathcal{X}}\rangle_\Phi &=\frac{\alpha^3\beta}{(1+q)^{1/2}M}\mathcal{P}_{\Phi,\mathcal{X}},
\\
\langle \dot{\mathcal{X}}\rangle_\Xi
&=\frac{\alpha^5\beta}{M}\mathcal{P}_{\Xi,\mathcal{X}},
\end{align}
with $\mathcal{X}\in\{\varphi_0,\phi_0,i\}$. Note that $\langle\dot e\rangle_\Xi=0$. Here $\mathcal{P}$ functions are nondimensional and generally depend on the orbital parameters $\{x_a,e,i,\varphi_0\}$. Through straightforward calculation, we obtain analytical approximations to the $\mathcal{P}$ functions of $|1011\rangle$ cloud valid to high orders of eccentricity, which are provided in the supplementary files. It should be stressed that we are treating the cloud's gravitational influence on the companion as a small perturbation and consider only its secular effects\footnote{We also neglect the short-term evolution of osculating orbital elements in evaluating the dissipative fluxes, such as $P_\text{gw}$ and $P_\text{ion}$.}, this approximation ceases to be valid if the cloud mass is sufficiently large. But in such cases the modifications on the orbital motion is also likely to be more distinguishable.

\begin{figure*}[htb]
	\centering
	\includegraphics[width=1\textwidth]{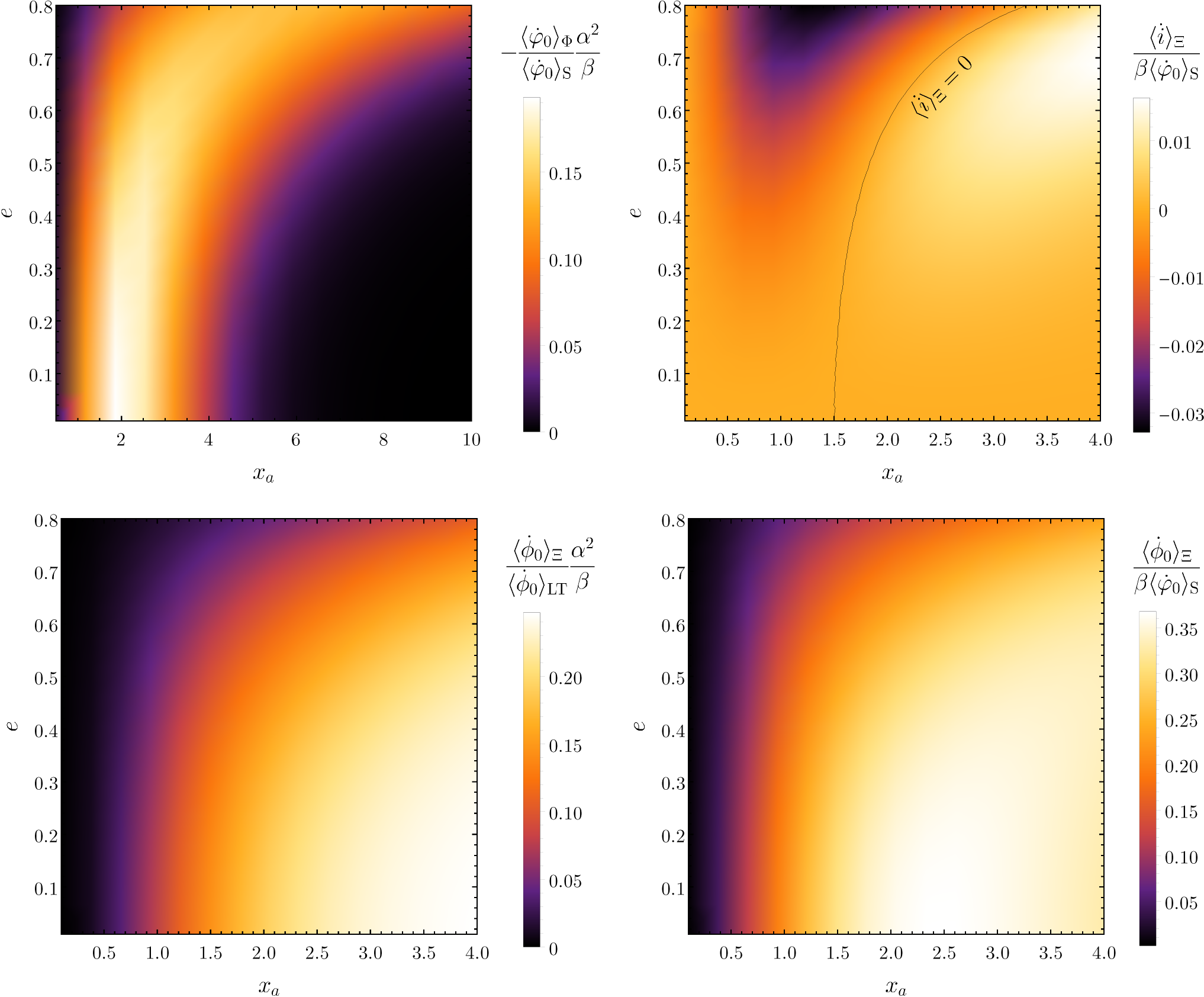}
	\caption{Precession rates of eccentric orbit in the $|1011\rangle$ cloud. For $\langle \dot{\mathcal{X}} \rangle_\Xi$ and $\langle \dot\phi_0 \rangle_\text{LT}$, we set $\varphi_0=\pi/4$, $i=\pi/2$ and $\chi=4\alpha$.}\label{precession_compare}
\end{figure*}

In the far field, the Newtonian potential sourced by the cloud can be approximated by $\Phi\approx -\frac{M_\text{c}}{r}+\frac{Q_\text{c}}{r^3}\frac{1-3\cos^2\theta}{2}$,  provided that the quadrupole moment scalar $Q_\text{c}=\int d^3r\,r^2\rho(\mathbf{r})\,P_2(\cos\theta)$ is nonzero \cite{Su:2021dwz,Cao:2023fyv}. In this approximation, the secular orbital precession rates are
\begin{align}
	\langle\dot\varphi_0\rangle_\Phi&=
	\frac{\alpha^3\beta}{(1+q)^{1/2}M}
	\frac{3(5 \cos 2i+3)}{8 \left(1-e^2\right)^2 x_a^{7/2}}\left(-\frac{Q_\text{c}}{M_\text{c}r_\text{c}^2}\right)
	,
	\\
	\langle \dot \phi_0 \rangle_\Phi&=
	\frac{\alpha^3\beta}{(1+q)^{1/2}M}
	\frac{-3\cos i}{2(1-e^2)^2x_a^{7/2}}\left(-\frac{Q_\text{c}}{M_\text{c}r_\text{c}^2}\right)
	,
\end{align}
with $\langle \dot i\rangle_\Phi=\langle \dot e\rangle_\Phi=0$. The quadrupole approximation breaks down at small radius; for a spherically symmetric mass distribution such as the vector $|1011\rangle$ cloud, $Q_\text{c}$ vanishes (neglecting the tidal deformation of the cloud \cite{Arana:2024kaz}), and the precession rates have to be computed more accurately. Note that $\langle \dot{\mathcal{X}}\rangle$ only captures the secular evolution of an osculating element $\mathcal{X}$ averaged over many orbital periods. This is illustrated in Fig.~\ref{precession_check} for the case of vector $|1011\rangle$ cloud, where the exact periastron shift and its secular approximation $\varphi_0(t)\approx\varphi_0(0)+\langle \dot\varphi_0\rangle t$ are compared. In this figure we also show the result taking into account the fast-oscillating term in the gravitational potential \eqref{Phi_full} and find it indeed negligible at such orbital radius.

In Fig.~\ref{precession_compare}, we plot the magnitudes of $\langle\dot{\mathcal{X}}\rangle_{\Phi,\Xi}$ relative to the BH-induced Schwarzschild and LT precession rates. The strongest effect of the cloud is a negative contribution to the periastron precession $\langle \dot\phi_0 \rangle_\Phi$, which is enhanced relative to $\langle\dot\varphi_0\rangle_\text{S}$ by a factor of $\beta/\alpha^2$, provided that $\beta\sim\alpha$. $\boldsymbol{\Xi}$ also contributes to the periastron precession (with $\langle \dot\varphi_0 \rangle_\Xi \propto \cos i$) but is $\alpha^2$-suppressed relative to $\langle \dot\varphi_0 \rangle_\Phi$ and is thus negligible. The cloud-induced orbital plane precession $\langle \dot\phi_0 \rangle_\Xi$ is always positive, with $\mathcal{P}_{\Xi,\phi_0}=A+B\cos 2\varphi_0$, where $A,B$ are functions of $\{x_a,e\}$ and $|B|\ll |A|$, so it is nearly $\varphi_0$-independent. The cloud leads to a change of the inclination angle as well (with $\langle \dot i \rangle_\Xi \propto \sin i\,\sin 2\varphi_0$), which is absent in the vacuum case, although the rate of change is typically much smaller than $\langle \dot\phi_0 \rangle_\Xi$.

\begin{figure}[t]
	\centering
	\includegraphics[width=0.45\textwidth]{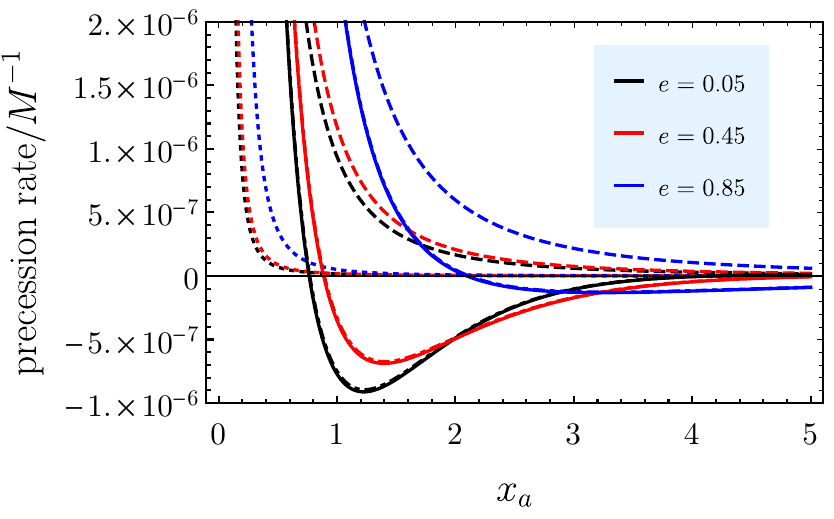}
	\caption{Comparison of orbital precession rates in the vector $|1011\rangle$ cloud, for $\alpha=\beta=0.05$. Dashed lines: $\langle \dot\varphi_0\rangle_\text{S}$, dot-dashed lines: $\langle \dot\varphi_0\rangle_\text{S}+\langle \dot\varphi_0\rangle_\Phi$, solid lines: $\langle \dot\varphi_0\rangle_\text{S}+\langle \dot\varphi_0\rangle_\Phi+\langle \dot\varphi_0\rangle_\Xi(\varphi_0=\pi/4, i=0)$, dotted lines: $\langle \dot\phi_0\rangle_\Xi(\varphi_0=\pi/4)+\langle \dot\phi_0\rangle_\text{LT}(\chi=4\alpha)$.}\label{precession_compare_2}
\end{figure}

Sufficiently far away from the cloud, the periastron precession is dominated by $\mathbf{F}_\text{1PN}$, while the orbital plane precession is dominated by the cloud's gravitomagnetic field with $\langle \dot\phi_0 \rangle_\Xi\approx\langle \dot\phi_0 \rangle_\text{LT}(\chi=\beta/\alpha)$. Both effects are however weak in this regime. If $\beta/\alpha^2$ is large enough, the periastron precession can be dominated by the cloud during an intermediate regime, which is overall shifted to larger radius when the eccentricity is higher (see the upper-left panel of Fig.~\ref{precession_compare} and also the solid lines in Fig.~\ref{precession_compare_2}). As the companion inspirals, the Schwarzschild precession finally dominates, while the precession of orbital plane becomes non-negligible at a still later stage, where it is dominated by the spin of the host BH. Thus even if the GA carries the same amount of angular momentum as the original spinning BH before the SR growth, the effect of spin precession is reduced.

\begin{figure*}[bth!]
	\centering
	\includegraphics[width=0.95\textwidth]{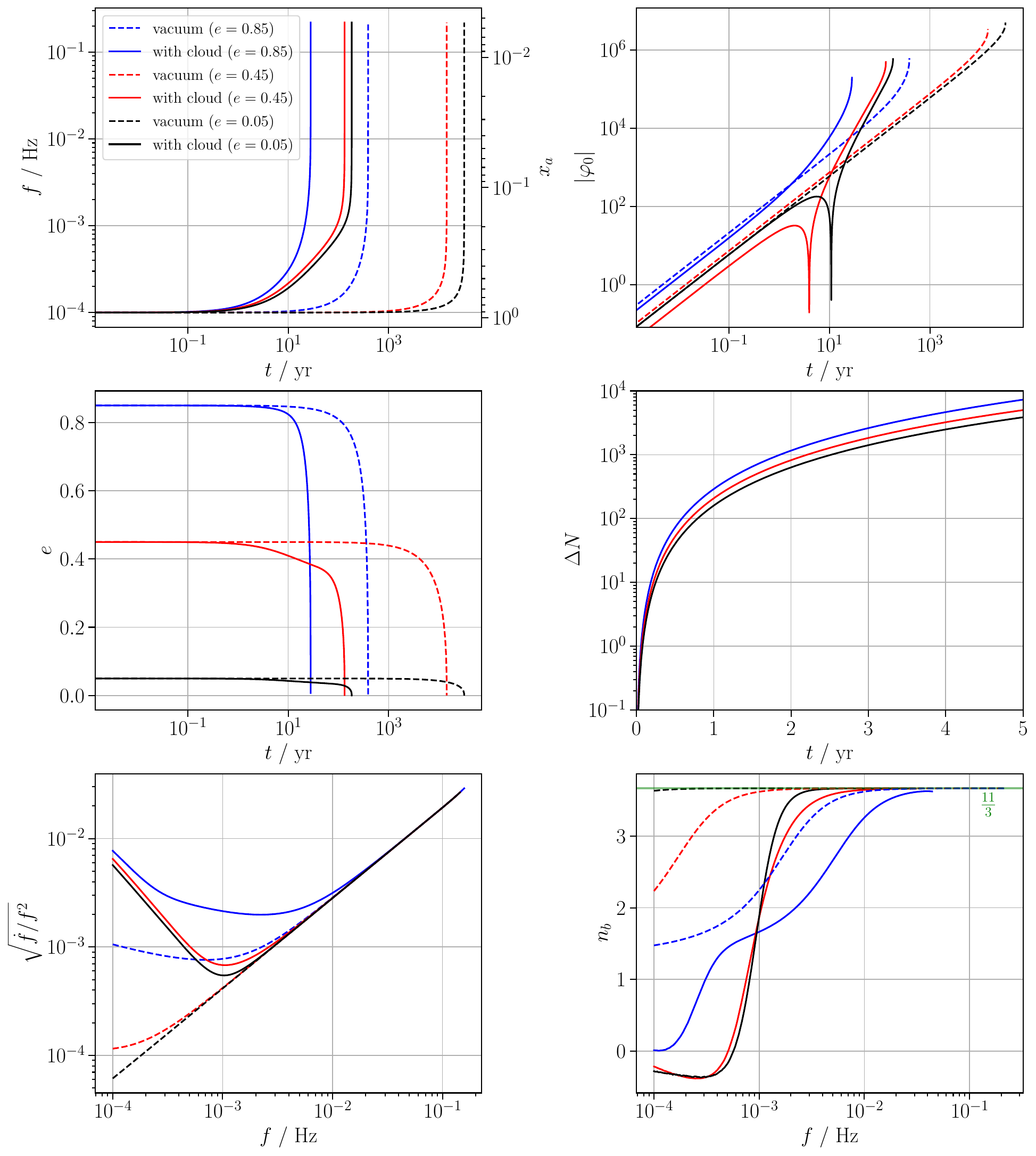}
	\caption{Adiabatic orbital evolution of IMRI in the vector $|1011\rangle$ cloud for $m_1=10^4M_\odot$, $m_2=10M_\odot$, $\alpha=0.03$, $\beta=0.8\alpha$, initial frequency $f=10^{-4}$ Hz (corresponding to $x_a\approx 0.9$) and different initial orbital eccentricities, shown are the temporal evolution of the orbital frequency $f$, the argument of periastron $|\varphi_0|$, the orbital eccentricity $e$ and the dephasing $\Delta N$ relative to a vacuum system with same initial condition, as well as the evolution of $(\dot f/f^2)^{1/2}$ and the braking index $n_b$ with $f$.}\label{orbital_evolution_plot}
\end{figure*}

\section{Gravitational waveforms of IMRIs in the vector $|1011\rangle$ cloud}\label{sec_3}
Using the perturbative model presented in the previous section, we can now concretely examine the inspirals in the vector $|1011\rangle$ cloud and the associated GW waveforms. Before considering specific systems, we first discuss some general aspects of the problem.

The companion can induce resonances during the inspiral. For the case of scalar GA in the most relevant $|211\rangle$ and $|322\rangle$ states, an inspiraling companion is likely to encounter adiabatic (hyper)fine resonances or strong nonresonant effects \cite{WY_3} at large orbital separation (though the orbital evolution may be accelerated due to a common envelope of the binary \cite{Guo:2024iye}). If this happens, only in some limited configurations can it cross the resonance bands without disrupting the cloud \cite{Tomaselli:2024bdd}. In contrast, inspiraling companions in the vector $|1011\rangle$ cloud can only induce sinking Bohr resonances and the backreaction from nonresonant transitions is estimated to be weak if the companion is sufficiently small (see Appendix~\ref{appendix_Bohr} for details). Both effects are therefore not expected to significantly change the state of the cloud and thus shall be neglected, since we focus on the off-resonant systems (or it can be assumed that the inspiral starts with $x_a<x^{(1)}$, which is the minimum semimajor axis of Bohr resonance). As discussed in Appendix~\ref{appendix_DC}, we can also neglect the cloud depletion effect.

The considered vector GA has SR origin. In this respect the major differences with the scalar GA are: (i) due to a stronger SR instability for the same boson mass and BH spin, the vector cloud can have a reasonably short growth timescale for smaller values of $\alpha$; (ii) due to a more rapid GW depletion rate for the same boson mass, also only relatively small values of $\alpha$ would be relevant if we assume a reasonably old cloud with $\beta/\alpha>0.1$. For a saturated cloud, $\alpha$ cannot be too small, in which case the cloud would take too much time to grow\footnote{The case of a growing cloud in the GA-companion system is also possible. If the growth is sufficiently slow, the cloud can be treated as a saturated one, otherwise there can be interplay between ionization and SR growth. The growth of the cloud will also lead to the evolution of the Newtonian potential $\Phi-M/r$ of GA, the effect of such processes is discussed in Appendix~\ref{appendix_DC}.}.

Assuming the SR growth of $|1011\rangle$ state saturates at the time $t\sim t_1$, the age of this cloud at $t=t_0$ is $\tau_\text{c}\approx t_0-t_1$. If the cloud depletion is solely due to $P_\text{gw,c}$, $M_\text{c}$ evolves from its initial saturated value $M_\text{c,0}$ according to $-\dot M_\text{c}=P_\text{gw,c}=\beta^2p(\alpha)$ (note the $p(\alpha)$ function is state-dependent, for the $|1011\rangle$ cloud we use the accurate result provided by Ref.~\cite{Siemonsen:2022yyf}), leading to $M_\text{c}(t)=M_\text{c,0}/[1+(t-t_1)/\tau_\text{gw}(t_1)]$, where we introduced the cloud depletion timescale $\tau_\text{gw}(t)\equiv M_\text{c}(t)/P_\text{gw,c}(t)=M/[\beta(t) p(\alpha)]$. Taking the initial BH mass and spin (before the SR process) to be $M_i$ and $\chi_i\sim 1$, the cloud mass at SR saturation is $M_{c,0}\approx \alpha M_i\chi_i/m\sim \alpha M/m$ with $m$ being the azimuthal quantum number of this state, the present-day cloud mass is then given by ${\beta_\text{crit}(\tau_\text{c})}/{\alpha}\equiv
1/(m+p\,\alpha\,\tau_\text{c}/M)$. The right panel of Fig.~\ref{1011_timescale} shows the relation between $\beta_\text{crit}$ and $\alpha$ for different BH masses. Note these constraints in principle do not exit for boson clouds with possible non-SR origins.

The typical situation for the dissipative orbital dynamics in the vector $|1011\rangle$ cloud is shown in the left panel of Fig.~\ref{1011_timescale} for the case of circular orbit. The depicted ionization power and mass flux are universal for $q\ll 1$, with $P_\text{gw}$ shifted upward if $\alpha^5/\beta$ becomes larger. Around its peak point $x^{(1)}\approx 1.6$, $P_\text{ion}$ can dominate over $P_\text{gw}$, which is the ideal regime to probe the cloud through the accelerated orbital decay. Note in this regime the orbital separation is much larger than the radius of innermost stable circular orbit (ISCO) and the radius at which $\Omega=\mu$ (depicted as the vertical red dotted and solid lines in the left panel of Fig.~\ref{1011_timescale}, respectively), justifying our neglect of the relativistic oscillation of the vector field.

In the absence of electromagnetic processes, gravitational waves from such a GA-companion system provide the only observable. We are mainly interested in the GW emitted by the binary (those giving rise to $P_\text{gw}$). For the computation of its waveform we use the analytical kludge (AK) approximation as reviewed in Appendix~\ref{appendix_GW}. For $q\ll 1$, the fundamental frequency of this GW is
\begin{equation}
\begin{aligned}
f\equiv \frac{\Omega}{2\pi}\approx\frac{10^4M_{\odot}}{M}\left(\frac{\alpha}{0.03}\right)^{3}\left(\frac{1}{x_a}\right)^{3/2}8\times 10^{-5}\mathrm{Hz},
\end{aligned}
\end{equation}
with $\alpha=0.1\times(\mu/1.3\times 10^{-11}\text{eV})(M/M_\odot)$. We can thus see that extreme mass-ratio inspirals around SMBH with $M>10^5M_\odot$ cannot probe the $x_a\sim 1$ regime in the LISA band $f>10^{-4}$ Hz if $\alpha\lesssim 0.07$.

The gravitational radiation of the cloud leads to another GW signal. In the case of vector $|1011\rangle$ cloud, it is given by $h_+=H_0 [(1+\cos^2\theta)/2]\cos\Upsilon$ and $h_\times=H_0 \cos\theta\sin\Upsilon$, where $H_0=\sqrt{10 P_\text{gw,c}}/\left[2\omega^{(1011)} d\right]$, $\Upsilon =2\omega^{(1011)} t -2\phi+\text{const}$, $\cos\theta=-\mathbf{e}_z\cdot\mathbf{n}$ ($\mathbf{n}$ is the unit vector pointing from the detector to the BH, $d$ is the source-detector distance) and the polarization tensors are defined by the spherical basis vectors $\mathbf{a}=\mathbf{e}_\theta$, $\mathbf{b}=\mathbf{e}_\phi$ in the $x'y'z'$ frame (see Appendix~\ref{appendix_GW}). Since $\omega^{(1011)}\approx \mu$, this is a nearly monochromatic signal at frequency much higher than $f$. The ratio between the time-domain strain amplitudes of  the binary GW and cloud GW is roughly measured by ${h_\text{binary}}/{h_\text{cloud}}\sim {h_0}/{H_0}\approx 4q\alpha^3/[\sqrt{10\,p}\,\beta(1-e^2)x_a]$. For the $|1011\rangle$ cloud this is proportional to $q/[\alpha^2\beta(1-e^2)x_a]$, the cloud GW is thus typically negligible for $x_a\sim 1$ if $q$ is not too small.

Based on the considerations above and for concreteness, we now focus on a specific set of binary parameters: $m_1=10^4M_\odot$, $m_2=10M_\odot$, $\alpha=0.03$, $\beta=0.813\,\alpha$, $d=1\,\text{Mpc}$, which corresponds to $\mu=4\times10^{-16}\,\text{eV}=0.6\,\text{Hz}$, $H_0=6.7\times 10^{-23}$ and the cloud age $\tau_\text{c}=10^6\,\text{yr}$ for $\chi_i=1$. Since we focus on the low-frequency systems, the minor effects of cloud-induced orbital plane precession and the inclination angle change (as well as the higher PN corrections) will be neglected, hence for the computation of GW waveforms, we also fix the parameters $\varphi(t=0)=0$, $B+\varphi_0(t=0)=\pi/2$, $\bar\phi_0=\bar\alpha_0=0$, $\theta_\text{L}=2\pi/3$, $\phi_\text{L}=3\pi/2$, $\theta_\text{S}=\pi/6$, $\phi_\text{S}=\pi/5$, where $(\theta_\text{S},\phi_\text{S})$ and $(\theta_\text{L},\phi_\text{L})$ are the spherical coordinates of $\mathbf{n}$ and $\mathbf{e}_Z$ (whose orientation depends on $\{\phi_0,i\}$) in the solar-system-barycenter (SSB) frame described in Appendix~\ref{appendix_GW}.

Figure~\ref{orbital_evolution_plot} plots the orbital evolution of this system starting with $f(t=0)=10^{-4}$ Hz and different orbital eccentricities. In the presence of $|1011\rangle$ cloud (solid lines), the decay rates of both the semimajor axis and eccentricity are greatly enhanced compared with the vacuum system (dashed lines). The ratio $(\dot f/f^2)^{1/2}$ however remains small, this ensures fast bound-free transitions and justifies the neglect of transient processes in the ionization model \cite{B4}. To measure the deviation from vacuum evolution, we introduce the accumulated equal-time dephasing $\Delta N\equiv \int_0^{t}[f(t')-f_\text{vac}(t')]dt'$. A larger initial eccentricity results in a faster inspiral and increased dephasing. This is simply due to the monotonic dependence of the ionization power on $e$. In addition to the frequency evolution, the GW phase is also affected by the periastron shift. For $e=0.05$ and $0.45$, the periastron initially undergoes negative precession (the sharp dip in the plot of $|\varphi_0(t)|$ is due to a change of sign, before the dip $\varphi_0$ is negative). However this is not observed for $e=0.85$, since in the present case (similar to the situation in Fig.~\ref{precession_compare_2}) $\langle  \dot\varphi_0\rangle_\Phi$ is suppressed relative to the Schwarzschild precession when the orbit is more eccentric. Had the inspiral started with an orbital separation $x_a>x^{(1)}$, the companion will also encounter ionization jump, during which there will be a sharp change in the chirping rate $\dot f$; for the considered systems this is however difficult to observe by the LISA-like detectors since it happens at a too low orbital frequency. We do not show the evolution of cloud mass under ionization, since in all cases the relative change of $\beta$ during the whole inspiral is about $10^{-2}$ (consistent with the assumption of perturbativity). The braking index $n_b\equiv f\ddot f/(\dot f)^2$ of a vacuum system is
\begin{equation}
n_b=\frac{4 \left(407 e^8+2344 e^6+93810 e^4+58720 e^2+25344\right)}{3 \left(37 e^4+292 e^2+96\right)^2}.
\end{equation}
It takes the minimum value $4/3$ at $e=1$ and the maximum value $11/3$ at $e=0$, the latter is eventually approached by all systems due to the orbital circularization and $P_\text{ion}$ being less significant. In the presence of cloud, smaller or even negative braking index can be realized during the inspiral.

Finally, we evaluate the detectability of the events as well as the distinguishability between the vacuum and nonvacuum waveforms. Given the time-domain waveform $h_{\lambda=\text{I,II}}(t)$ in $t\in[t_\text{min},t_\text{max}]$, the frequency-domain waveform is obtained as the Fourier transform $\tilde h_\lambda(f)\equiv \int_{t_\text{min}}^{t_\text{max}}dt\,h_\lambda(t)\,e^{2\pi if t}$. The inner product between two waveforms is defined by
\begin{equation}
\begin{aligned}
\langle h_1|h_2 \rangle \equiv  2\sum_{\lambda=\text{I},\text{II}}\int_{f_\text{min}}^{f_\text{max}} df\,\frac{\tilde h_{1,\lambda}(f)\,\tilde h_{2,\lambda}^*(f)+\text{c.c.}}{S_n(f)}
,
\end{aligned}
\end{equation}
where $S_n$ is the single-sided noise power spectrum density of the detector. For reference, we consider the LISA detector with $f_\text{min}=10^{-4}\,\text{Hz}$ and $f_\text{max}<1\,\text{Hz}$. The optimal signal-to-noise ratio (SNR) of a waveform $h_\text{I,II}(t)$ is $\text{SNR}^2=\langle h|h \rangle=\int_{f_\text{min}}^{f_\text{max}} d \ln f\,\left({h_c}/{\sqrt{fS_n}}\right)^2$, with $h_c(f)\equiv 2f\sqrt{|\tilde h_\text{I}(f)|^2+|\tilde h_\text{II}(f)|^2}$ being the characteristic strain. The overlap between two waveforms can be measured by the faithfulness $\mathcal{F}\equiv\langle h_1|h_2 \rangle/\sqrt{\langle h_1|h_1 \rangle \langle h_2|h_2 \rangle}$, and a rough estimation for the condition of distinguishability is \cite{PhysRevD.95.104004} $\text{SNR}>\text{SNR}_\text{crit}\equiv\sqrt{D/[2(1-\mathcal{F})]}$, where $D$ is the number of parameters in the waveform template. In the present model, a vacuum system has $D=11$.

\begin{figure*}[thb!]
	\centering
	\includegraphics[height=0.393\textwidth]{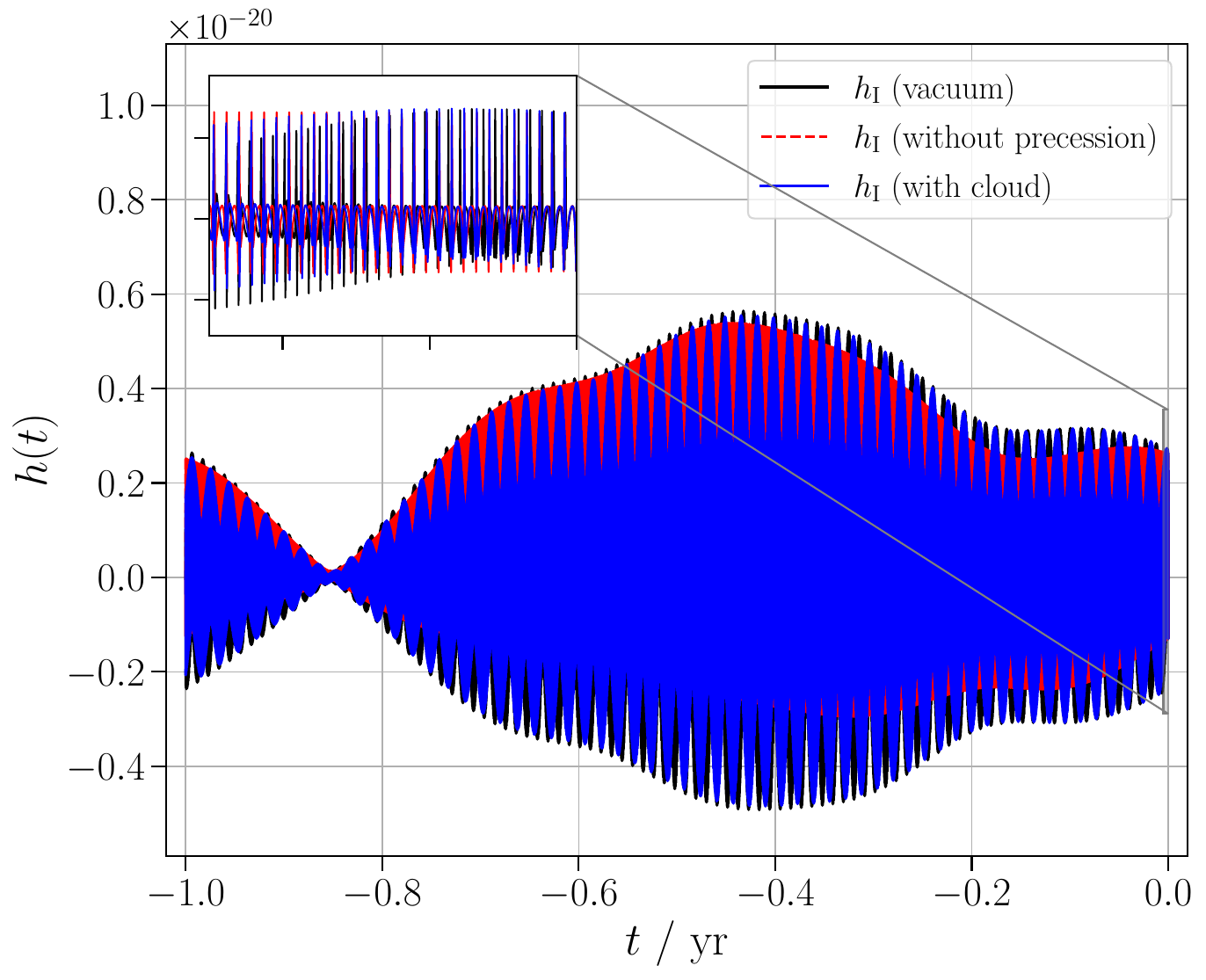}
	\,
	\includegraphics[height=0.385\textwidth]{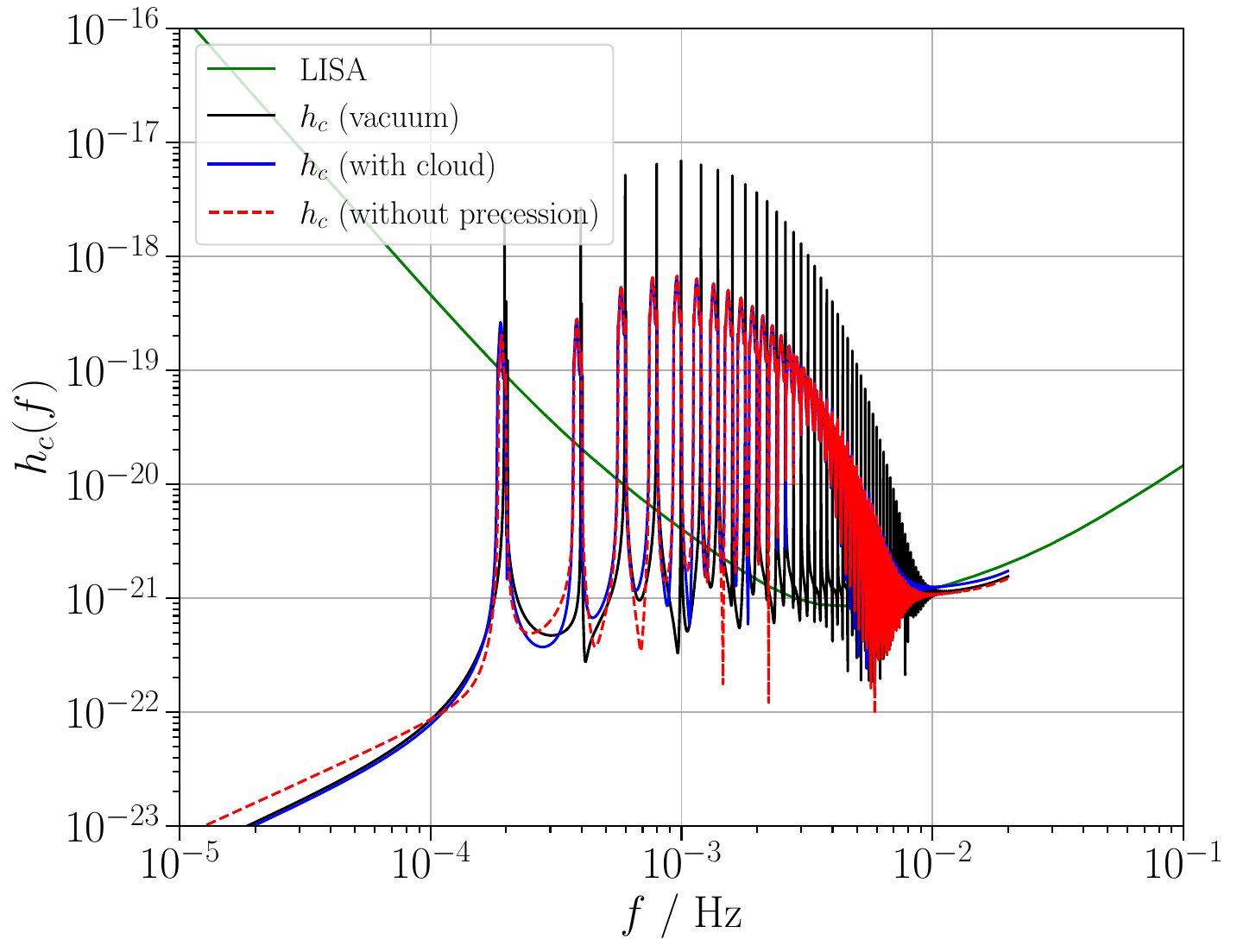}
	\\
	\,
	\includegraphics[height=0.393\textwidth]{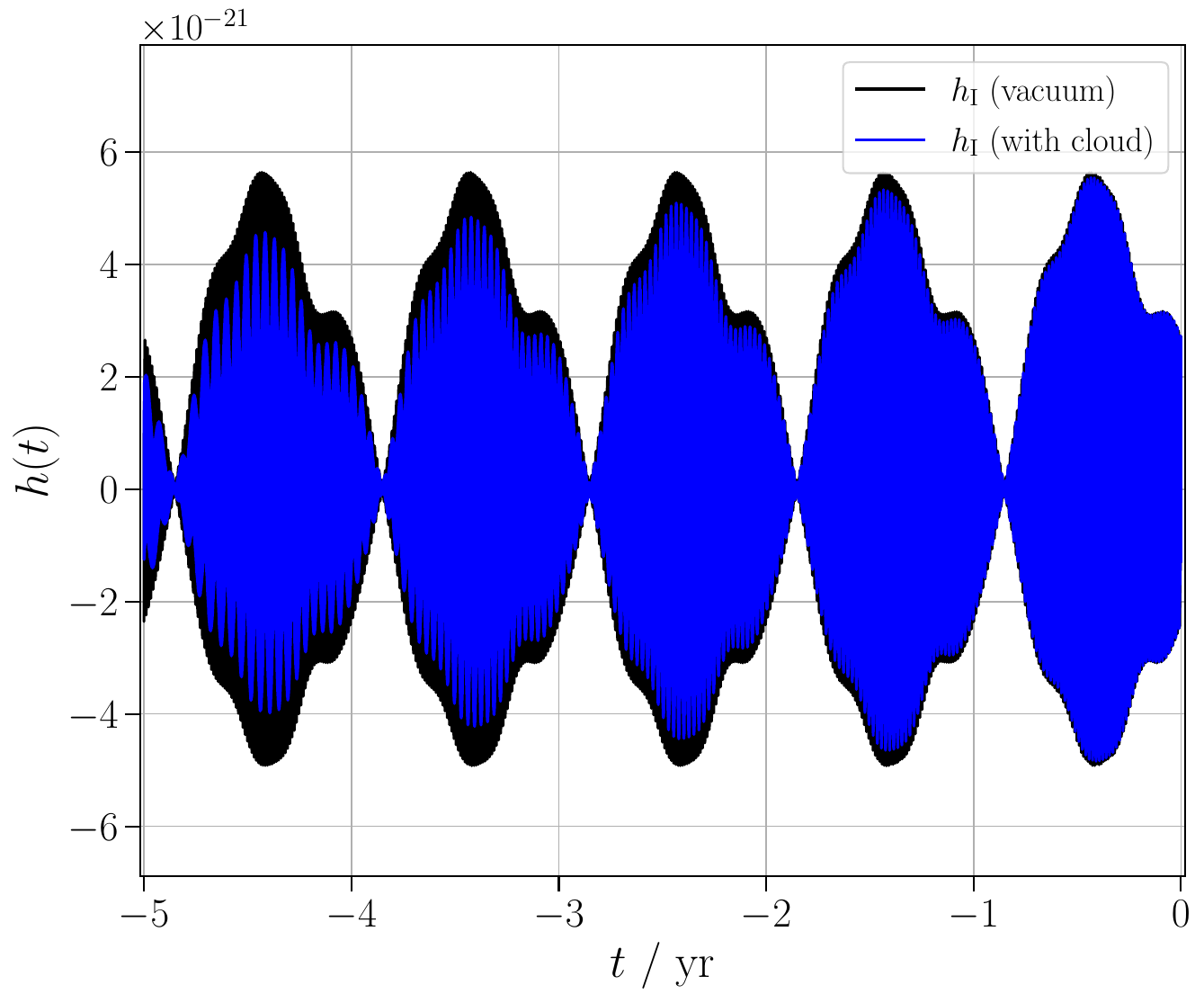}
	\,
	\includegraphics[height=0.385\textwidth]{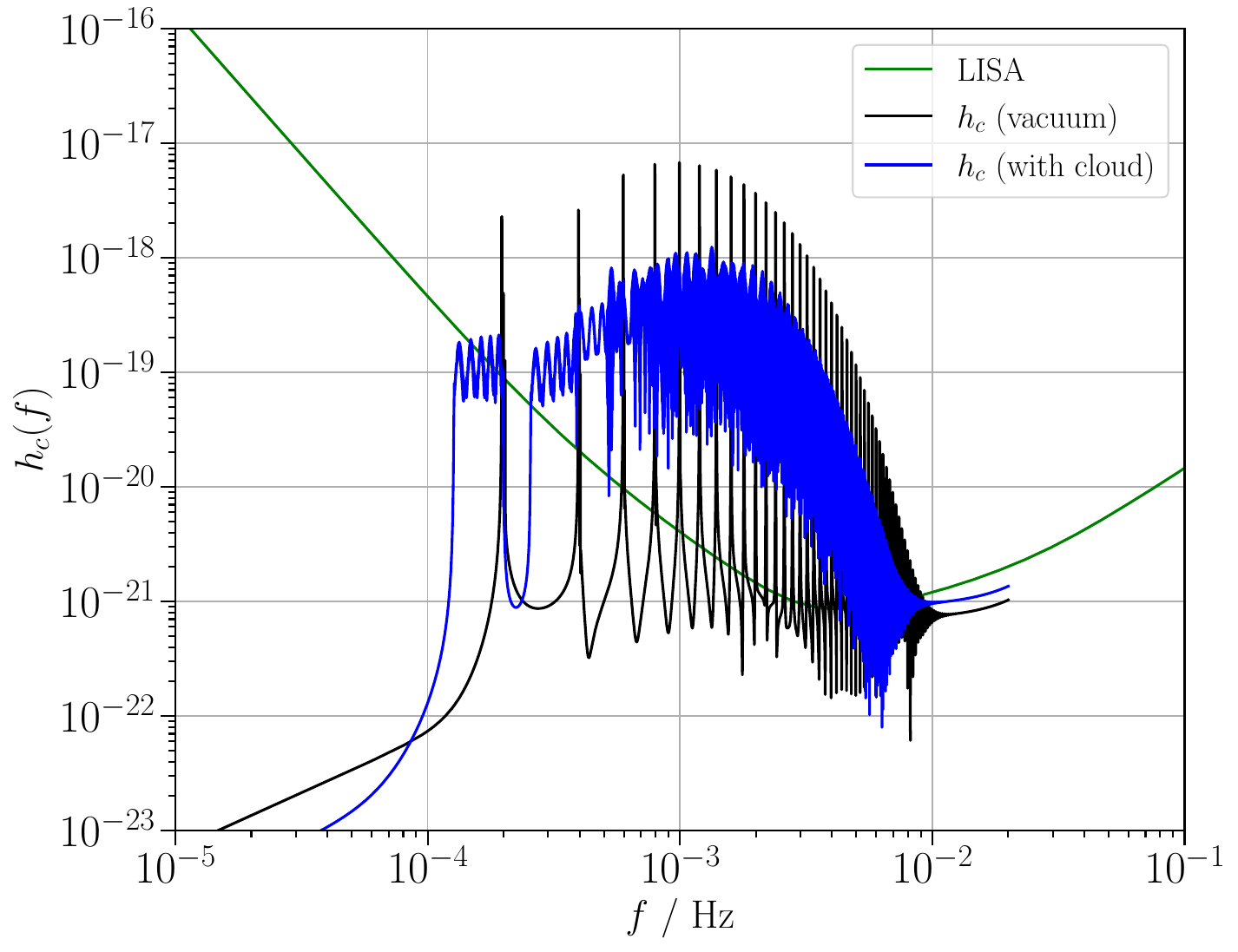}
	\caption{One-year (upper row) and five-year (lower row) time-domain waveform $h_\text{I}(t)$ and frequency-domain characteristic strain spectrum $h_c(f)$ (shown against the noise curve $\sqrt{f S_n}$) of IMRIs with $m_1=10^4M_\odot$, $m_2=10M_\odot$, $\alpha=0.03$, $\beta=0.8\alpha$ and given final state $\{f_0=2\times 10^{-4}\text{Hz},e_0=0.6\}$, in the presence (blue line) or absence (black line) of $|1011\rangle$ cloud. The optimal SNRs and overlaps between the vacuum and nonvacuum waveforms are $\text{SNR}(\text{1 yr})=114$, $\text{SNR}_\text{vac}(\text{1 yr})=124$, $\mathcal{F}(\text{1 yr})=-10^{-3}$; $\text{SNR}(\text{5 yr})=189$, $\text{SNR}_\text{vac}(\text{5 yr})=289$, $\mathcal{F}(\text{5 yr})=-9\times10^{-4}$. We also show the one-year waveform taking into account the cloud ionization but without including the periastron precession (red line), whose optimal SNR is 116. The annual amplitude modulation of the time-domain waveform is due to the time-dependent response function of LISA. Note the one-year optimal SNR of the cloud GW $\sim H_0\sqrt{ 1\,\text{yr}/S_n(f=\mu/\pi)}=4.35$.}\label{waveform_1}
\end{figure*}

We choose to compare the waveforms of vacuum and nonvacuum systems with orbital parameters matched at $t_\text{max}=0$, with $f(t=0)= f_0$, $e(t=0)= e_0$ and $\beta(t=0)=\beta$. The one-year and five-year waveforms for $f_0=2\times 10^{-4}\,\text{Hz}$ and $e_0=0.6$ are plotted in Fig.~\ref{waveform_1}. The amplitude and phase modulations from periastron precession are visible in the time domain, as can be seen from a comparison with the waveform of constant $\varphi_0$ (depicted in the red line). A five-year observation already reveals the secular amplitude evolution due to the accelerated orbital decay in the presence of cloud, in contrast to the nearly invariant amplitude in the vacuum case. The accelerated frequency evolution manifests in the frequency spectrum as the broadening and lowering of harmonic peaks. Since the periastron precession rate ($\sim 10^{-5}$ Hz) is considerably smaller than $\Omega$, the splits of harmonic peaks are not observed in the frequency spectrum. The inclusion of orbital precession is however necessary due to its non-negligible contribution to the GW phase. For the one-year waveforms in Fig.~\ref{waveform_1}, the overlap between waveforms with and without precession is only about 0.01. Also note that due to the scaling property of orbital evolution, for given $\{x_a(t=0),e_0,q,\alpha,\beta,d\}$, the time-domain waveform can be written as $h(t)=M\mathcal{H}(t/M)$, where $\mathcal{H}(z)$ is a universal function, the inner product though cannot be simply rescaled since $S_n$ is frequency-dependent. At given orbital frequency, the rates of change of $e$ and $x_a$ (as well as SNR) are proportional to the mass ratio $q$, while the periastron precession rate is $q$-insensitive.

\begin{figure*}[thb!]
	\centering
	\includegraphics[width=0.48\textwidth]{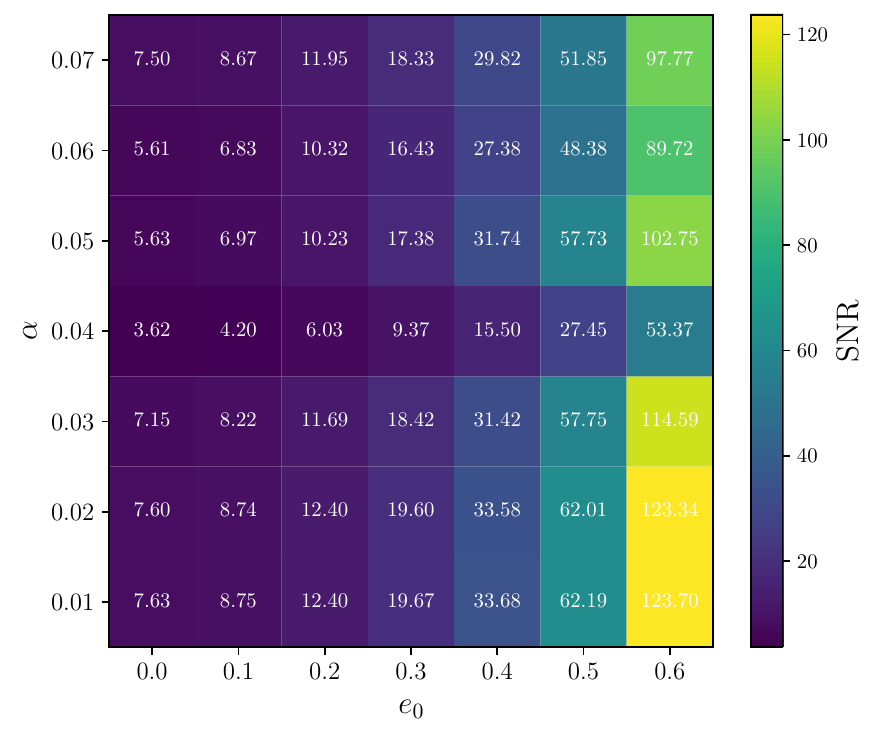}
	\includegraphics[width=0.48\textwidth]{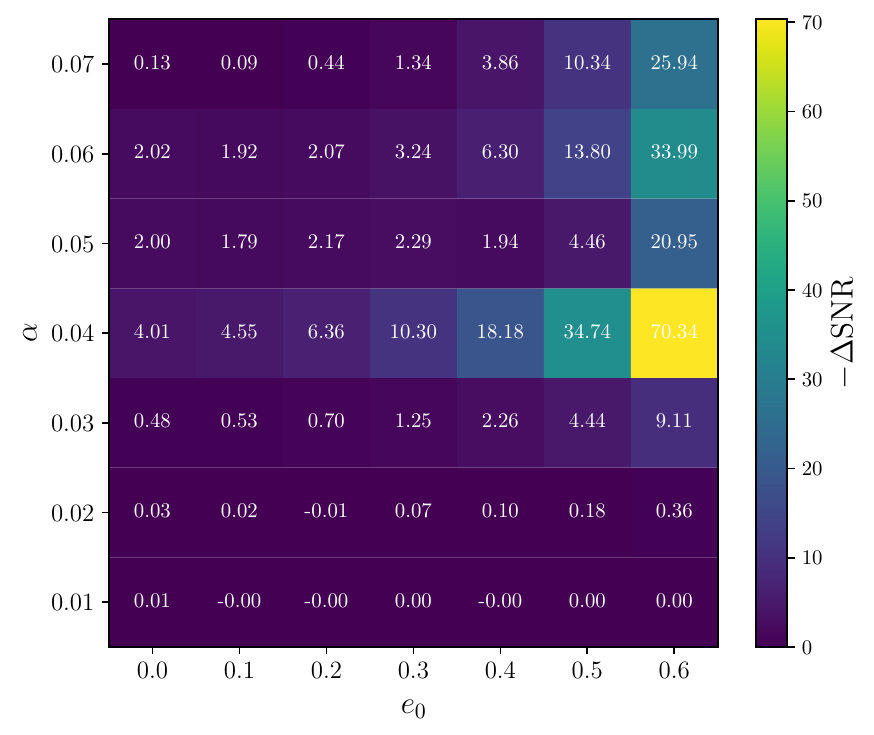}
	\includegraphics[width=0.48\textwidth]{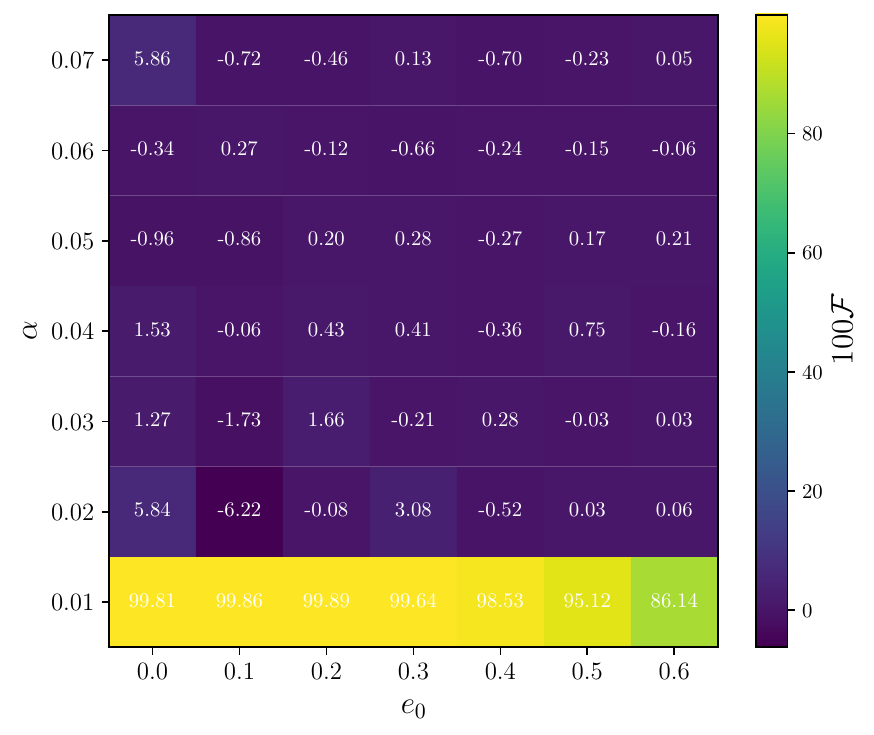}
	\includegraphics[width=0.48\textwidth]{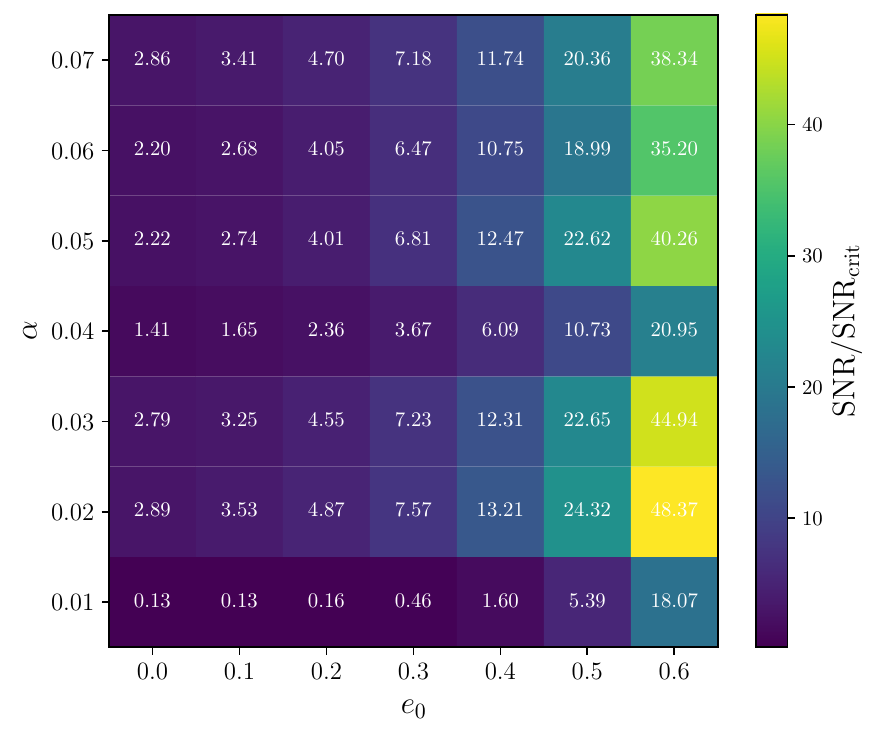}
	\includegraphics[width=0.48\textwidth]{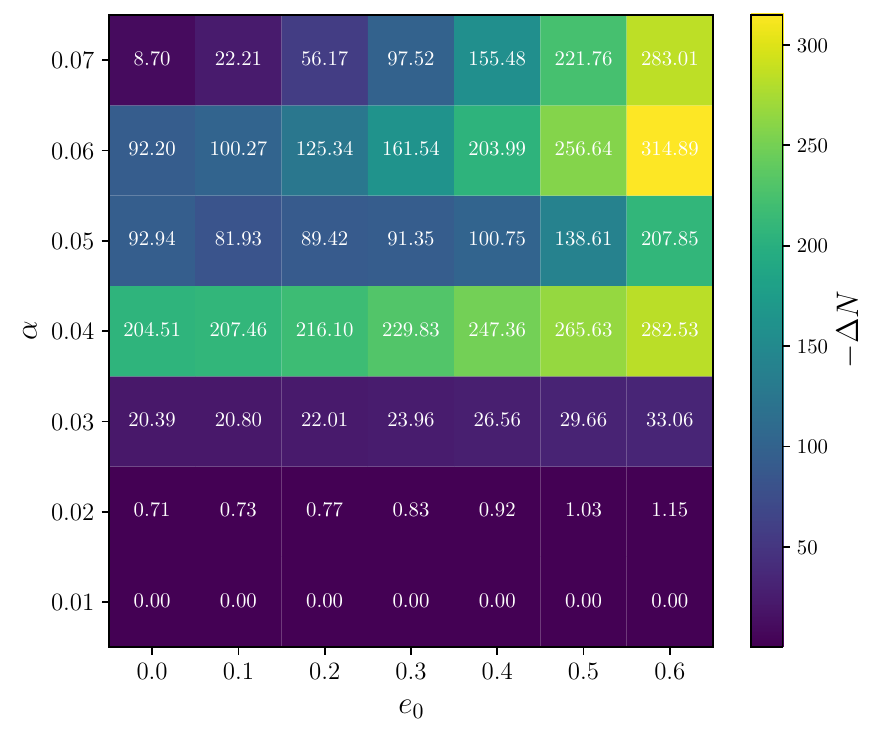}
	\includegraphics[width=0.48\textwidth]{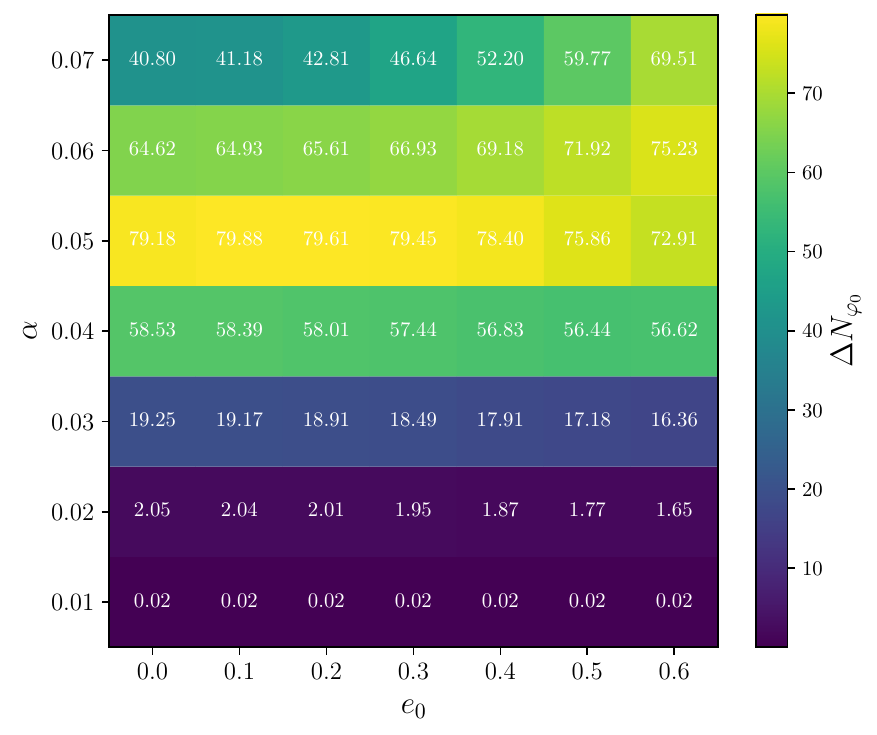}
	\caption{Comparison between the one-year vacuum and nonvacuum waveforms for different values of $\alpha$ and final eccentricity $e_0$, with $\beta/\alpha=0.813$ and the setting of other parameters is same as Fig.~\ref{waveform_1}. Shown are the optimal SNR in the presence of cloud (upper left panel), $\Delta\text{SNR}=\text{SNR}-\text{SNR}_\text{vac}$ (upper right panel), the overlap $\mathcal{F}$ between vacuum and nonvacuum waveforms (middle left panel), $\text{SNR}/\text{SNR}_\text{crit}$ for $D=11+2$ (middle right panel), the dephasing $\Delta N$ (lower left panel) and the periastron shift difference $\Delta N_{\varphi_0}$ (lower right panel).}\label{waveform_2}
\end{figure*}

In Fig.~\ref{waveform_2} we display the one-year waveforms with varying $\alpha,e_0$ and constant $\beta/\alpha$. For $\alpha> 0.01$, the faithfulness $\mathcal{F}$ is around zero; for $\alpha=0.01$, the effect of cloud becomes weak due to the excessively large Bohr radius. The periastron shift difference $\Delta N_{\varphi_0}\equiv -\frac{1}{2\pi}\int_{-1\,\text{yr}}^{0}[\langle \dot\varphi_0\rangle(t')-\langle \dot\varphi_0\rangle_\text{vac}(t')]\,dt'$ appears to be more sensitive to $\alpha$, although the orbital evolution at given frequency generally does not depend on $\alpha$ in a simple way (see Appendix~\ref{appendix_given-frequency} for more discussions). The optimal SNR is enhanced for larger eccentricity and also the cloud becomes more distinguishable, as reflected on the increase of SNR difference, dephasing $\Delta N\equiv \int_{-1\,\text{yr}}^{0}[f(t')-f_\text{vac}(t')]\,dt'$ and the ratio $\text{SNR}/\text{SNR}_\text{crit}$.

\begin{figure*}[htb!]
	\centering
	\includegraphics[width=1.05\textwidth]{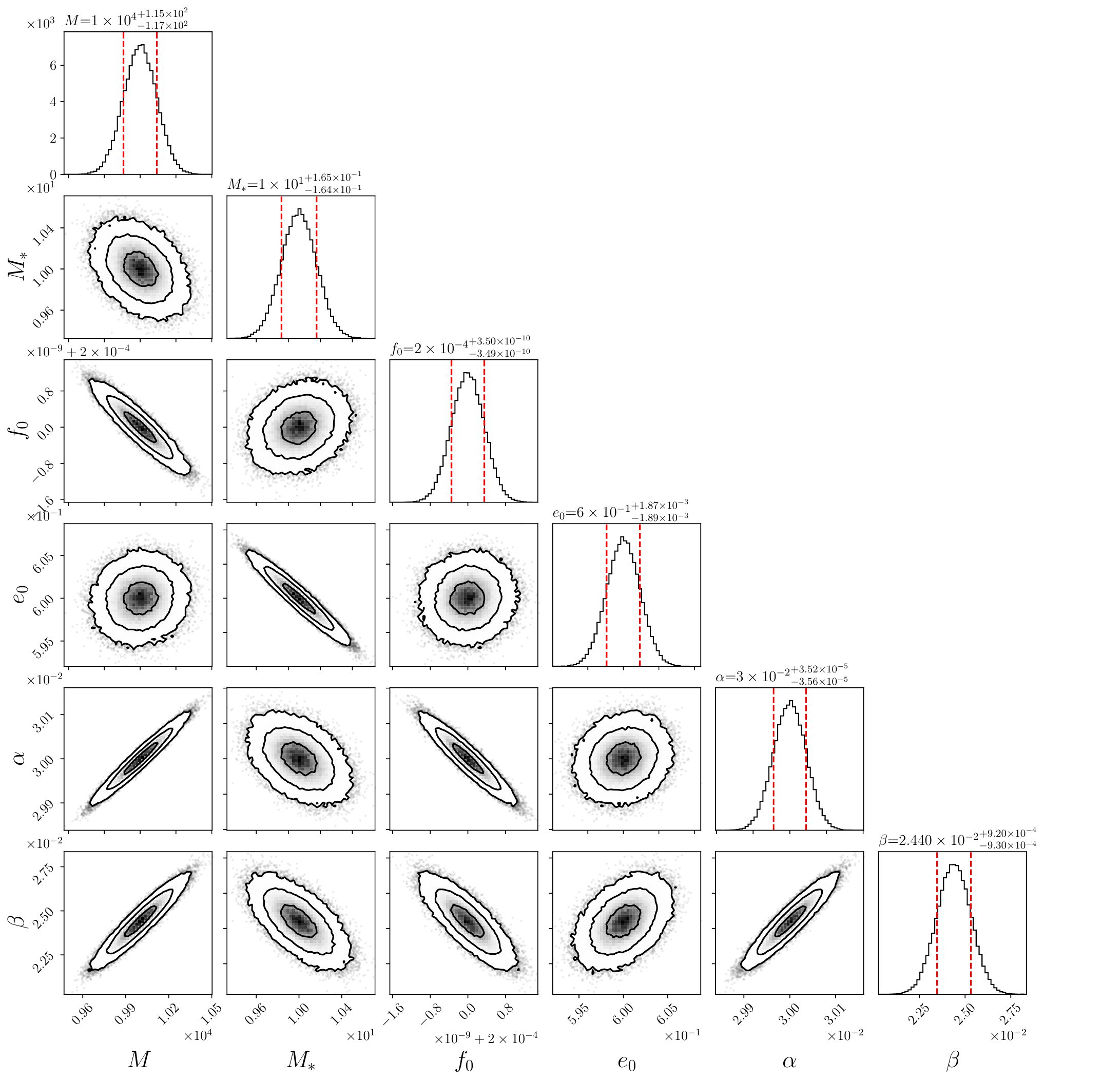}
	\caption{Corner plot depicting the probability distribution of $\{M/M_\odot,M_*/M_\odot,f_0/\text{Hz},e_0,\alpha,\beta\}$ inferred from one-year observation of LISA as estimated by the FIM. The contours mark the $\{1,2,3\}\sigma$ levels of the 2D distributions, and the red dashed lines represent the $\{16\%,84\%\}$ quantiles of the marginalized distributions. The $1\sigma$ statistical errors obtained from the FIM are $\sigma_M/\hat M=1.2\times 10^{-2}$, $\sigma_{M_*}/\hat M_*=1.6\times 10^{-2}$, $\sigma_{f_0}/\hat f_0=1.8\times 10^{-6}$, $\sigma_{e_0}/\hat e_0=3.1\times 10^{-3}$, $\sigma_\alpha/\hat\alpha=1.2\times 10^{-3}$ and $\sigma_\beta/\hat\beta=3.8\times 10^{-2}$. Note $\sigma_{f_0}$ is compatible with the estimation \cite{Seto:2001pg} $(\text{SNR}\times\text{1\,yr})^{-1}\sim 3\times10^{-10}\,\text{Hz}$. For simplicity, other parameters $\{d,\theta_\text{S},\phi_\text{S},\theta_\text{L},\phi_\text{L},B+\varphi_0(t=0),\varphi(t=0)\}$ are fixed. The statistical errors estimated by the FIM are proportional to the source distance $d$, which is chosen to be 1 Mpc.}\label{Fisher}
\end{figure*}

The detection accuracy can be estimated using the Fisher information matrix (FIM). For a model $h_{\boldsymbol{\xi}}(t)$ with parameters $\boldsymbol{\xi}=\{\xi_i\}_i$, the FIM evaluated at the true parameter values $\boldsymbol{\xi}=\boldsymbol{\hat \xi}$ is $\Gamma_{ij}=\left\langle {\partial h}/{\partial \xi_i}\right.\left|\, {\partial h}/{\partial \xi_j}\right\rangle_{\boldsymbol{\xi}=\boldsymbol{\hat \xi} }$, and the inferred covariance matrix of $\boldsymbol{\xi}$ is estimated to be $\Sigma_{ij}=(\Gamma^{-1})_{ij}$, with the root-mean-squared error of $\xi_i$ given by $\sigma_i=\sqrt{\Sigma_{ii}}$. We consider a six-parameter model with $\boldsymbol{\xi}=\{M,M_*,f_0,e_0,\alpha,\beta\}$, $\boldsymbol{\hat \xi}=\{10^4\,M_\odot,10\,M_\odot,2\times 10^{-4}\,\text{Hz},0.6,0.03,0.0244\}$ and the observation time span $t\in[-1,0]\,\text{yr}$, the corner plot of the resulting probability distribution is shown in Fig.~\ref{Fisher}. The obtained $1\,\sigma$ relative errors of the cloud parameters $\{\alpha,\beta\}$ are smaller than unity, showing enough precision to identify the presence of the cloud. Meanwhile, $\sigma_\beta$ is considerably larger than the fractional change of $\beta$ during the observation time span, which is about $6\times 10^{-4}$, so taking $\beta$ to be constant can be a good approximation. We also calculated the FIM for $\hat\beta=0$ with $\alpha$ fixed to $0.03$ and the true values of other parameters unchanged, the estimated $1\,\sigma$ constraint for $\beta$ is given by $\beta<\sigma_{\beta}=3\times 10^{-6}$. This could effectively rule out the existence of such a vector boson, if we assume a finite age $\tau_\text{c} < 10^{10}\,\text{yr}$ for the saturated cloud.

\section{Summary and discussion}\label{sec_4}
Ultralight vector field with suitable particle mass can be spontaneously produced from the spinning BHs via SR processes, forming vector GAs. We have investigated the off-resonant orbital dynamics of a small companion around the vector GA occupied by the fastest-growing $|1011\rangle$ state, taking into account the backreaction of cloud ionization and the conservative perturbations from cloud gravity. The cloud-induced effects and the leading relativistic corrections are compared in both the dissipative and conservative sectors. The model has a simple scaling symmetry and is thus quite general for $q\ll 1$. However, the validity of Newtonian treatment and the cloud's SR origin demands a reasonably small value of $\alpha$ if the mass of the cloud is not too small. We have therefore focused on the low-frequency IMRI systems with $f\sim 10^{-4}$ Hz, for which the imprints of cloud on the orbital evolution and hence the GW waveform of the binary through accelerated orbital decay and modified periastron precession are more distinguishable. As is demonstrated in Sec.~\ref{sec_3}, detection of such a binary as well as the cloud can be possible with the next-generation space-borne GW detectors, especially if the system has a high eccentricity.

Similar models can be constructed for other vector GA states, e.g., the second fastest-growing state $|2122\rangle$. Though for the systems we considered, its growth rate is too slow to be relevant. The scalar field decomposition of $|2122\rangle$ is $\boldsymbol{\Psi}_{2122}=(1/\sqrt{2})(\Psi_{21-1},-i\Psi_{211} ,0)^T$, where $\Psi_{21,\pm 1}$ is the wave function of scalar $|21,\pm1\rangle$ state. The ionization fluxes of $|211\rangle$ cloud in the cases of equatorial orbit and inclined circular orbit have been calculated in Ref.~\cite{Tomaselli:2023ysb}, for a general inclined eccentric orbit the computation would be more difficult (as discussed in Appendix~\ref{appendix_ionization}). Similar to the case of scalar $|211\rangle$ cloud, there are also issues of low-frequency resonances for the vector $|2122\rangle$ cloud, which would require a dedicated study.

Although we have focused on vector GA, the ionization fluxes and the stationary Newtonian potential $\Phi$ of the vector $|1011\rangle$ state are degenerate with those of scalar $|100\rangle$, vector $|101m'\rangle$ and tensor $|102m'\rangle$ states (in principle, the different $m'$-states are distinguishable by their different gravitomagnetic fields). The same model can thus be applied to the nonrelativistic scalar $|100\rangle$ cloud formed through spherical accretion or quartic self-interaction \cite{Budker:2023sex}. In such scenarios, the presence of orbiting companions around a central body can possibly impose stronger constraints on such a scalar halo due to its forced ionization.

\section*{Acknowledgment}
Y.C. thanks Mu-Chun Chen and Yue-Hui Yao for useful discussions. Y.T. is supported by the National Key Research and Development Program of China (No. 2021YFC2201901).

\cleardoublepage
\twocolumngrid

\appendix
\section{Metric perturbations of the cloud}\label{appendix_metric perturbations}
Far away from its host BH, the linear metric perturbations sourced by the cloud can be approximately computed in a flat spacetime background \cite{Ferreira:2017pth,Chen:2022kzv}, with
\begin{equation}
	ds^2=-(1+2\Psi)dt^2+2\Xi_idx^idt+\left[(1-2\Phi)\delta_{ij}+h_{ij}\right]dx^idx^j,\label{perturbed_metric}
\end{equation}
we choose the gauge $\partial_i \Xi_i=h_{ii}=\partial^ih_{ij}=0$. For a given matter source $T_{ab}$, the linearized Einstein equation can be written as \cite{maggiore2018gravitational}
\begin{align}
	\nabla^2\Phi&=4\pi \rho,\label{Einstein_1}
	\\
	\ddot\Phi+\frac{1}{3}\nabla^2\Psi&=\frac{4\pi}{3} (\rho+3p),\label{Einstein_2}
	\\
	4\dot \Phi_{,i}-\nabla^2 \Xi_i&=16\pi T_{i0},\label{Einstein_3}
	\\
	(\Phi-\Psi)_{,ij}-2\dot \Xi_{(i,j)}-\square h_{ij}&=16\pi T_{\langle ij \rangle}.\label{Einstein_4}
\end{align}
here $\rho=T_{00}$, $p=\frac{1}{3}T_{ii}$, and we denote $X_{\langle ij \rangle}\equiv X_{ij}-\frac{1}{3}\delta_{ij}X_{kk}$, $X_{(i,j)}\equiv\frac{1}{2}(X_{i,j}+X_{j,i})$. We focus on the vector $|1011\rangle$ state, which corresponds to the vector field profile:
\begin{equation}\label{1011_field}
	\frac{\mathbf{A}(t,\mathbf{r})}{\alpha^2\beta^{1/2}}\approx
	-\frac{e^{-x}}{\sqrt{\pi}}
	\left(\begin{matrix}
		\cos \omega t
		\\
		\sin \omega t
		\\
		0
	\end{matrix}\right),\quad \partial_tA_0\approx \nabla\cdot\mathbf{A},
\end{equation}
where $\omega=\omega^{(1011)}\sim \mu$. The energy-momentum tensor is\footnote{Since Eq.~\eqref{1011_field} is not a free field solution in the flat spacetime, $T_{\mu\nu}$ evaluated with $g_{ab}\approx\eta_{ab}$ does not satisfy $\partial^a T_{ab}=0$, hence we expect that only the leading-$\alpha$ results for $\Psi$ and $\boldsymbol{\Xi}$ are reliable.}
\begin{equation}
	T_{a b}=-\frac{1}{4} g_{a b}F_{cd}F^{cd}+{F_a}^d F_{b d}+\mu^2\left(A_a A_b-\frac{1}{2} g_{a b} A_cA^c\right).
\end{equation}
The induced metric perturbations oscillate at frequency $2\omega$. However, for sufficiently large semimajor axis $x_a\gg\alpha^{2/3}(1+q)^{1/3}$ the oscillating perturbations effectively average out within one orbital period\footnote{For the same reason, the perturbing force and vector radiation \cite{PhysRevD.49.6892,2024arXiv240100204C,Zi:2024mbd} arising from the possible nongravitational couplings of the companion with the vector field will also be irrelevant.}, while at small enough radius where such oscillation is non-negligible, the effect of the cloud typically becomes much weaker. Therefore in the main text we considered only the stationary part of the leading-order scalar perturbation given by Eqs.~\eqref{Einstein_1} and \eqref{Einstein_2}:
\begin{align}
		\frac{\Psi}{M_\text{c}/r_\text{c}}&=
		-\frac{1}{x}+e^{-2 x} \left(\frac{1}{x}+1\right)\nonumber
		\\
		&\quad +\frac{9}{2}\left[
		\frac{1}{x^3}-\frac{1}{3x}+e^{-2 x} \left(-\frac{1}{x^3}-\frac{2}{x^2}-\frac{5}{3x}-\frac{2}{3}\right)
		\right]\nonumber
		\\
		&\quad\quad\times\sin^2\theta\,\cos(2\omega t-2\phi)+\mathcal{O}(\alpha^2) \label{Phi_full}
		\\
		&\approx
		-\frac{1}{x}+e^{-2 x} \left(\frac{1}{x}+1\right)
		\approx \frac{\Phi}{M_\text{c}/r_\text{c}}.\label{Phi}
\end{align}
This is simply the Newtonian potential sourced by the nonrelativistic mass distribution $\rho_{1011}=\frac{M_\text{c}}{r_\text{c}^3}\frac{e^{-2x}}{\pi}$.

Using also Eq.~\eqref{Einstein_3} we find the stationary part of the vector perturbation sourced by the cloud: $\boldsymbol{\Xi}=-\frac{2 J_\text{c}(\mathbf{r}) \sin \theta}{r^2}\mathbf{e}_{\phi}$, with the effective spin profile:
\begin{equation}
	J_\text{c}(\mathbf{r})/(M_\text{c}/\mu)=e^{-2 x} \left(-2 x^2-2 x-1\right)+1.\label{J_c}
\end{equation}
In the far field limit, $J_\text{c}\to M_\text{c}/\mu$, which is precisely the angular momentum carried by the $|1011\rangle$ cloud. Also notice that Eq.~\eqref{J_c} coincides with the profile of enclosed mass $\tilde\beta(x)=x^2\partial_x\left(\frac{\Phi}{M_\text{c}/r_\text{c}}\right)$ given by Eq.~\eqref{Phi}, and is compatible with the uniform NR angular momentum density per unit mass given by Eq.~\eqref{j}. Finally, using Eq.~\eqref{Einstein_4} we can obtain the cloud's leading-order GW radiation power\footnote{This computation of gravitational radiation in the flat spacetime approximation gives the correct $\alpha$-scaling for $\alpha \ll 1$ but underestimates the GW flux for a saturated cloud \cite{Siemonsen:2022yyf}.} \cite{Baryakhtar:2017ngi} $P_\text{gw,c}\approx \beta^2\left(\frac{32}{5}\alpha^{10}\right)$ and the associated GW waveform as presented in Sec.~\ref{sec_3}.

In the background \eqref{perturbed_metric}, the geodesic equation (up to linear order in both the velocity $\mathbf{v}$ and metric perturbations) takes the form $\dot v_i\approx-\partial_i\Psi-\partial_t \Xi_i+[\mathbf{v}\times\left(\nabla\times\boldsymbol{\Xi}\right)]_i-v^j\partial_t\left[h_{ij}-(2\Phi+\Psi)\delta_{ij}\right]$, hence the dominant effect on a test body comes from the gravitoelectric potential $\Psi$, and the next correction is provided by the gravitomagnetic potential $\boldsymbol{\Xi}$.

In the presence of stationary scalar and vector perturbations\footnote{Note the cloud may also be perturbed by a GW background \cite{Liu:2024mzw}.} (with $\Phi\approx\Psi$), the wave equation of $A_k$ reads
\begin{equation}
	\begin{aligned}
		&\Big[\delta_{kl}(\square+2\Phi\partial_t^2+2\Phi\nabla^2+2\Xi_i\partial_i\partial_t)
		\\
	    &-(\partial_l \Xi_k)\partial_t
		+(\partial_k \Xi_l)\partial_t
		\\ 
		&
		+2\delta_{kl}(\partial_i\Phi)\partial_i
		-2(\partial_l\Phi)\partial_k
		\Big]A_l
		\approx\mu^2 A_k,
	\end{aligned}
\end{equation}
where we kept only terms linear in $\{\Phi,\boldsymbol{\Xi}\}$ and neglected the spatial gradients of $A_0$. Correspondingly the nonrelativistic dynamics of $\boldsymbol{\Psi}$ defined in Eq.~\eqref{wave function} is\footnote{In comparison, the nonrelativistic wave equation of a scalar field $\phi=\frac{1}{\sqrt{2\mu}}(\psi e^{-i\mu t}+\text{c.c.})$ is $i\partial_t\psi=-\frac{1}{2\mu}\nabla^2\psi+i\Xi_l\partial_l\psi+\mu\Phi\psi$.}
\begin{equation}\label{NR_vector_EOM}
	\begin{aligned}
		i\partial_t\Psi_k
		=&
		-\frac{1}{2\mu}\nabla^2\Psi_k+\mu\Phi\Psi_k
		\\
		&+i\,\Xi_l\partial_l\Psi_k-\frac{i}{2}(\partial_l \Xi_k)\Psi_l
		+\frac{i}{2}(\partial_k \Xi_l)\Psi_l
		\\
		&
		-\frac{1}{\mu}(\partial_l\Phi)\partial_l\Psi_k
		+\frac{1}{\mu}(\partial_l\Phi)\partial_k\Psi_l
		.
	\end{aligned}
\end{equation}
The last two terms in the second line comes from the coupling between the gravitomagnetic field and the spin angular momentum density $\mathbf{j}_\text{spin}$ in Eq.~\eqref{j}. Including the energy $-\tilde{\mathbf{j}}\cdot \boldsymbol{\Xi} - \left(\frac{1}{2}\mathbf{j}_\text{spin}\right)\cdot \left(\nabla\times\boldsymbol{\Xi}\right)$, with $\tilde{\mathbf{j}}=(M_\text{c}/\mu)\big[-\frac{i}{2}(\Psi_k^*\nabla\Psi_k-\text{c.c.})\big]$, Eq.~\eqref{epsilon} becomes $\epsilon =(M_\text{c}/\mu)\big[ \frac{1}{2\mu}\nabla\Psi_k\cdot\nabla \Psi_k^*+\mu\Phi \Psi_k\Psi_k^* + \frac{i}{2}(\Psi_k^*\Xi_l\partial_l\Psi_k-\text{c.c.})-\frac{i}{2}\Psi_l \Psi_k^* (\partial_l \Xi_k-\partial_k \Xi_l)\big]$, which leads to the first and second lines of Eq.~\eqref{NR_vector_EOM}.

But Eq.~\eqref{NR_vector_EOM} appears to be not accurate enough to describe the hyperfine structure of vector GA. The energy level correction $\langle nljm|H_\text{spin}|nljm\rangle=\Delta \omega^{(nljm)}_\text{spin}/q_j$ it gives (using $[H_\text{spin}\boldsymbol{\Psi}]_k=i\Xi_l\partial_l\Psi_k-\frac{i}{2}(\partial_l \Xi_k)\Psi_l
+\frac{i}{2}(\partial_k \Xi_l)\Psi_l$ and taking $\boldsymbol{\Xi}=-\frac{2M^2\chi}{r^2}\sin\theta\,\mathbf{e}_\phi$ for the BH spin) differs from the result $\Delta \omega^{(nljm)}_\text{spin}=\frac{16m\mu \chi \alpha^5}{(l+j)(l+j+1)(l+j+2)}$ of Ref.~\cite{B2} by a fudge factor $q_{j=l+1}=l/j$ and $q_{j\ne l+1}=(l+1)/(j+1)$.

\section{Quadrupole gravitational waveform}\label{appendix_GW}
The leading-order GW waveform of a binary is given by the quadrupole formula
\begin{equation}
\begin{aligned}
		h_{ij}^\text{TT}(t) &=\frac{2}{d}\Lambda_{ij,kl}(-\mathbf{n})\,\ddot{Q}_{kl}(t-d)\\
        &=e_{ij}^+\, h_+(t)+e_{ij}^\times \, h_\times(t),
\end{aligned}
\end{equation}
\begin{widetext}
\noindent where $d$ is the binary's distance to the detector (neglecting the cosmological redshift), $Q_{ij}$ the trace-free part of the quadrupole tensor $M_{ij}\equiv\nu M_\text{tot}\,(\mathbf{r}_*)_i(\mathbf{r}_*)_j$ evaluated in the center-of-mass frame, and $\Lambda_{ij,kl}=P_{ik}P_{jl}-\frac{1}{2}P_{ij}P_{kl}$ with $P_{ij}=\delta_{ij}-n_in_j$ the transverse-traceless (TT) projection operator for GW traveling in the $-\mathbf{n}$-direction. Here we neglect the Doppler modulation of GW frequency due to the binary's center-of-mass motion relative to the detector. The ortho-normalized polarization tensors can be constructed as $e_{ij}^+=a_ia_j-b_ib_j$ and $e_{ij}^\times=a_ib_j+a_jb_i$, with the three unit vectors $\{\mathbf{a,b,-n}\}$ forming a right-handed triad.

Specialized to a Keplerian binary with orbital normal $\mathbf{e}_Z$, choosing $\mathbf{a}=(\mathbf{n}\times\mathbf{e}_Z)/|\mathbf{n}\times\mathbf{e}_Z|$ and $\mathbf{b}=\mathbf{a}\times\mathbf{n}$, the quadrupole GW waveform in the time domain reads \cite{10.1093/mnras/274.1.115,Barack:2003fp}
\begin{equation}
		\begin{aligned}\label{waveform}
			\frac{h_+}{h_0}&=\left(1+c^2_\iota\right)\left[
			\left(\frac{5e}{4}c_\varphi+c_{2\varphi}+\frac{e}{4}c_{3\varphi}+\frac{e^{2}}{2}\right)
			c_{2B+2\varphi_0}
			+
			\left(\frac{5e}{4}s_\varphi+s_{2\varphi}+\frac{e}{4}s_{3\varphi}\right)
			s_{2B+2\varphi_0}
			\right]
			-
			s^2_\iota
			\left(\frac{e^2}{2}+\frac{e}{2}c_\varphi\right),
			\\
			\frac{h_\times}{h_0}&=c_\iota\left[
			\left(\frac{5e}{2}s_\varphi+2s_{2\varphi}+\frac{e}{2}s_{3\varphi}\right)
			c_{2B+2\varphi_0}-
			\left(\frac{5e}{2}c_\varphi+2c_{2\varphi}+\frac{e}{2}c_{3\varphi}+e^2\right)
			s_{2B+2\varphi_0}
			\right],
		\end{aligned}
\end{equation}
where $c_z\equiv \cos z$, $s_z\equiv \sin z$, $h_0\equiv \frac{2M M_*}{a(1-e^2)d}$, $\cos \iota\equiv -\mathbf{e}_Z\cdot\mathbf{n}$, and $B$ is the rotation angle from  $-[\mathbf{n}-(\mathbf{n}\cdot\mathbf{e}_Z)\,\mathbf{e}_Z]$ to $\mathbf{e}_Z\times\mathbf{e}_z$ about the positive $Z$-axis, with
\begin{align}
	\sin B  = \frac{(\mathbf{e}_Z\cdot\mathbf{n})\cos i-\mathbf{e}_z\cdot\mathbf{n}}{\sin i\sqrt{1-(\mathbf{e}_Z \cdot\mathbf{n})^2}} 
	,\quad
	\cos B  = \frac{\mathbf{n}\cdot(\mathbf{e}_z\times\mathbf{e}_Z)}{\sin i\sqrt{1-(\mathbf{e}_Z\cdot\mathbf{n})^2}}. 
\end{align}
For an adiabatically evolving osculating orbit, substituting the secularly varying orbital element $\mathcal{X}(t)$ (obtained from $\dot{\mathcal{X}}=\langle \dot{\mathcal{X}}\rangle$) into Eq.~\eqref{waveform} and computing $\varphi(t)$ with $\dot\varphi=\Omega(t)\frac{[1+e(t)\cos \varphi]^2}{\{1-[e(t)]^2\}^{3/2}}$ gives the AK waveform\footnote{Note the original AK model used the Fourier decomposition of \eqref{waveform}.} \cite{Seto:2001pg,Barack:2003fp,Mikoczi:2012qy}. In the case of a fixed orbital plane, the waveform depends solely on $\{\varphi(t),\varphi_0(t),a(t),e(t),B,\iota\}$.

The three arms of LISA form an orthogonal pair of two-arm detectors, in the low frequency approximation the two outputting signals are \cite{Cutler:1997ta}
\begin{equation}
h_\lambda(t)=\frac{\sqrt{3}}{2}\left[F^+_\lambda(t)\, h_+(t)+F^\times_\lambda(t)\, h_\times(t)\right],
\quad \lambda\in\{\text{I},\text{II}\},
\end{equation}
with the antenna pattern functions
\begin{align}
&F^+_\text{I}=\frac{1}{2}\left(1+c^2_{\theta_\text{s}}\right)c_{2\phi_\text{s}}c_{2\psi}
-c_{\theta_\text{s}} s_{2\phi_\text{s}} s_{2\psi},
&&F^\times_\text{I}=\frac{1}{2}\left(1+c^2_{\theta_\text{s}}\right)c_{2\phi_\text{s}}s_{2\psi}
+c_{\theta_\text{s}}s_{2\phi_\text{s}}c_{2\psi},
\\
&F^+_\text{II}=\frac{1}{2}\left(1+c^2_{\theta_\text{s}}\right)s_{2\phi_\text{s}}c_{2\psi}
+c_{\theta_\text{s}}c_{2\phi_\text{s}}s_{2\psi},
&&F^\times_\text{II}=\frac{1}{2}\left(1+c^2_{\theta_\text{s}}\right)s_{2\phi_\text{s}}s_{2\psi}
-c_{\theta_\text{s}}c_{2\phi_\text{s}}c_{2\psi},
\end{align}
and
\begin{align}
c_{\theta_\text{s}}=&\frac{1}{2}c_{\theta_\text{S}}-\frac{\sqrt{3}}{2}s_{\theta_\text{S}}\,c_{\bar\phi-\phi_\text{S}},
\\
\phi_\text{s}=&\bar \alpha_0+2\pi t/T_\text{d}
+\arctan \frac{\sqrt{3}c_{\theta_\text{S}}
+s_{\theta_\text{S}}c_{\bar\phi-\phi_\text{S}}}{2s_{\theta_\text{S}}s_{\bar\phi-\phi_\text{S}}},
\\
\tan \psi=&
\left[c_{\theta_\text{L}}-\sqrt{3}s_{\theta_\text{L}}c_{\bar\phi-\phi_\text{L}}
-2c_{\theta_\text{s}}(c_{\theta_\text{L}}c_{\theta_\text{S}}+s_{\theta_\text{L}}s_{\theta_\text{S}}c_{\phi_\text{L}-\phi_\text{S}})\right]
\nonumber
\\
&\times\left[s_{\theta_\text{L}}s_{\theta_\text{S}}s_{\phi_\text{L}-\phi_\text{S}}-\sqrt{3}\,c_{\bar\phi}(c_{\theta_\text{L}}s_{\theta_\text{S}}s_{\phi_\text{S}}-c_{\theta_\text{S}}s_{\theta_\text{L}}s_{\phi_\text{L}})
-
\sqrt{3}\,s_{\bar\phi}(c_{\theta_\text{S}}s_{\theta_\text{L}}c_{\phi_\text{L}}-c_{\theta_\text{L}}s_{\theta_\text{S}}c_{\phi_\text{S}})\right]^{-1}
,
\end{align}
with $\bar\phi=\bar\phi_0
+2\pi t/T_\text{d}$ and $T_\text{d}=1\,\text{yr}$. Here $(\theta_\text{S},\phi_\text{S})$ and $(\theta_\text{L},\phi_\text{L})$ are the spherical coordinates of $\mathbf{n}$ and $\mathbf{e}_Z$ in the fixed ecliptic-based $\tilde x\tilde y\tilde z$ frame centered at the solar system barycenter. $\mathbf{e}_Z$ can be obtained from $\mathbf{e}_z$ by a rotation about $\mathbf{e}_z\times\mathbf{e}_{\tilde z}$ by the angle $i$, followed by a rotation about $\mathbf{e}_z$ by an angle differing from $\phi_0$ only by a constant. In the absence of orbital plane precession and inclination angle change, $\{\theta_\text{L},\phi_\text{L},B\}$ simply stay constant.

For the noise spectrum of LISA, we use \cite{Robson_2019}
\begin{equation}
	S_n(f)=\frac{10}{3 L^2}\left\{P_{\mathrm{oms}}+2\left[1+\cos ^2\left(\frac{f}{f_*}\right)\right] \frac{P_{\mathrm{acc}}}{(2 \pi f)^4}\right\}\left[1+\frac{6}{10}\left(\frac{f}{f_*}\right)^2\right],
\end{equation}
with the optical metrology noise $P_{\mathrm{oms}}=\left(1.5 \times 10^{-11} \mathrm{~m}\right)^2\left[1+\left(2\, \mathrm{mHz}/f\right)^4\right] \mathrm{Hz}^{-1}$, the single test mass acceleration noise $P_{\mathrm{acc}}=\left(3 \times 10^{-15} \mathrm{~m} \mathrm{~s}^{-2}\right)^2\left[1+\left(0.4\, \mathrm{mHz}/f\right)^2\right]\left[1+\left(f/8\, \mathrm{mHz}\right)^4\right] \mathrm{Hz}^{-1}$, the arm length $L=2.5\,\text{Gm}$ and $f_*=19.09\,\text{mHz}$. Here we neglect the galactic confusion noise.
\end{widetext}

\section{Companion-induced level mixings}\label{appendix_mixings}
As a leading-order approximation to the orbital dynamics, both relativistic corrections and the cloud-induced effects can be neglected, the acceleration of the host BH is
\begin{equation}
	\ddot{\mathbf{R}}_1=\frac{M_*}{r_*^3}\mathbf{r}_*.
\end{equation}
Under Galilean transformation, this leads to an inertial acceleration in the BH-centered frame, which translates (under a suitable unitary transformation) to an effective linear potential for the NR wave function:
\begin{equation}
	\Phi_\text{inert}(t,\mathbf{r})=\frac{M_*}{r_*^3}\mathbf{r}_*\cdot\mathbf{r}.
\end{equation}
Note the coordinate in the static background frame is given by $\mathbf{r}'=\mathbf{r}+\mathbf{R}_1(t)$. The tidal gravitational potential experienced by the boson cloud in the BH-centered frame is thus
\begin{equation}\label{V_*}
	\Phi_*(t,\mathbf{r})=-\frac{M_*}{|\mathbf{r}_*(t)-\mathbf{r}|}+\Phi_\text{inert}(t,\mathbf{r}).
\end{equation}
Using the Laplace expansion:
\begin{equation}
	\frac{1}{|\mathbf{r}_1-\mathbf{r}_2|}=
	\sum_{l=0}^\infty \sum_{m=-l}^l
	\frac{4\pi}{2l+1}\frac{r_<^l}{r_>^{l+1}}Y^*_{lm}(\theta_1,\phi_1)Y_{lm}(\theta_2,\phi_2),
\end{equation}
where $r_>\equiv \text{max}(r_1,r_2)$ and $r_<\equiv \text{min}(r_1,r_2)$, the tidal potential energy $V_*=\mu \Phi_*$ can be expanded as
\begin{equation}
\begin{aligned}
V_*=\sum_{l_*=0}^\infty\sum_{m_*=-l_*}^{l_*} \frac{-4\pi q\alpha}{2l_*+1}Y_{l_*m_*}^*(\theta_*,\phi_*)Y_{l_*m_*}(\theta,\phi)\frac{F_{l_*}(x)}{r_\text{c}},
\end{aligned}
\end{equation}
with the radial function \cite{ZJ,Tomaselli:2023ysb}
\begin{equation}\label{F_function}
	F_{l_*}(x)=\begin{cases}
		\frac{x^{l_*}}{x_*^{l_*+1}}\Theta(x_*-x)+\frac{x_*^{l_*}}{x^{l_*+1}}\Theta(x-x_*),
		& l_*\ne 1
		\\
		\left(\frac{x_*}{x^2}-\frac{x}{x_*^2}\right)\Theta(x-x_*), & l_*=1
	\end{cases}
\end{equation}
where $\Theta(x)$ is the Heaviside unit step function.

We can now obtain the level-mixings of GA. The transition matrix element between two bound states $|i\rangle$ and $|i'\rangle$ is given by
\begin{equation}\label{bound-state-transition}
	\begin{aligned}
		\langle i'|V_*|i\rangle
		=\sum_{l_*,m_*}\frac{-4\pi q\alpha Y_{l_*,m_*}^*(\theta_*,\phi_*)}{2l_*+1}\frac{I_r}{r_\text{c}}I_\Omega \, e^{i(\omega'-\omega)t},
	\end{aligned}
\end{equation}
with the radial and angular integrals:\footnote{In the case of scalar GA, the vector spherical harmonics is replaced by the scalar one, with $I_\Omega=\int_0^\pi \sin\theta\,d\theta \int_0^{2\pi}d\phi\,Y_i(\theta, \phi)\,Y_{i'}^*(\theta, \phi)\,Y_{l_*m_*}(\theta,\phi)$ being the Gaunt coefficient.}
\begin{align}
	I_r&=\int_0^\infty dx\,x^2\,R_{i'}(x)F_{l_*}(x)R_{i}(x)
	,\label{I_r}
	\\
	I_\Omega&=\int_0^\pi \sin\theta\,d\theta \int_0^{2\pi}d\phi\,\mathbf{Y}_{i}(\theta, \phi)\cdot\mathbf{Y}_{i'}^*(\theta, \phi)\,Y_{l_*m_*}(\theta,\phi),\label{I_Omega}
\end{align}
where $i$ refers schematically to the set of relevant quantum numbers. Likewise the transition matrix element between a bound states $|i\rangle$ and a free state $|k\rangle$ is given by
\begin{equation}\label{bound-free}
	\begin{aligned}
		\langle k|V_*|i\rangle
		=\sum_{l_*,m_*}\frac{-4\pi q\alpha Y_{l_*,m_*}^*(\theta_*,\phi_*)}{2l_*+1}\frac{I_r}{\sqrt{r_\text{c}}}I_\Omega e^{i(\omega'-\omega)t},
	\end{aligned}
\end{equation}
where $I_r$ and $I_\Omega$ take similar forms, i.e., replacing $R_{i'}(x)$ with $R_{\tilde k}(x)$ in Eq.~\eqref{I_r}.

For a generic inclined eccentric orbit, apart from the factor $e^{i(\omega'-\omega)t}$, the time-dependence of the tidal mixing is also encoded in $I_r(x_*(t))$ and the spherical harmonics $Y_{l_*m_*}(\theta_*(t),\phi_*(t))$, the latter can be explicitly written as
\begin{equation}\label{rotation_of_Y}
	\begin{aligned}
Y_{l_*m_*}(\theta_*,\phi_*)=\sum_{g=-l_*}^{l_*}d^{l_*}_{m_*g}(-i)\,Y_{l_*g}\left(\frac{\pi}{2},0\right)\,e^{ig(\varphi_0+\varphi)},
	\end{aligned}
\end{equation}
here the Wigner $d$-matrix is defined by
\begin{equation}
	\begin{aligned}
		d^l_{m'm}(i)=&\sqrt{(l+m)!(l-m)!(l+m')!(l-m')!}\\&\times\sum_{k=k_\text{min}}^{k_\text{max}}(-1)^{k+m'-m} \frac{\left(\sin\frac{i}{2}\right)^{2k+m'-m}}{k!(l+m-k)!}
		\\
		&\quad \times
		\frac{\left(\cos\frac{i}{2}\right)^{2l-2k-m'+m}}{(l-m'-k)!(m'-m+k)!}
		~,
	\end{aligned}
\end{equation} 
with $k_\text{min}=\text{max}(0,m-m')$ and $k_\text{max}=\text{min}(l+m,l-m')$. Since the Wigner $d$-matrix, $Y_{l_*g}(\pi/2,0)$, $I_r$ and $I_\Omega$ are all purely real, the complex phase factor of a transition matrix element comes solely from $e^{i[g(\varphi_0+\varphi)+(\omega'-\omega)t]}$. For $|nlm\rangle=|100\rangle$, we can set $i=0$, the summation in Eq.~\eqref{rotation_of_Y} then involves only $m_*\in\{-l_*,-l_*+2,\cdots,l_*-2,l_*\}$ for which $Y_{l_*m_*}(\pi/2,0)$ is nonvanishing. we find that the contribution from $l_*\ge |m_*|+2$ is generally negligible, hence we include only the leading multipoles $l_*=|m_*|$ in our computation.

For the radial integral in the transition matrix element, the integration over $[x_*,\infty)$ can be conveniently removed as follows. For $l_*\ne 1$,
\begin{equation}
	\begin{aligned}
		I_r=
		\int_0^{x_*} dx\,x^2\,R'(x)\left(\frac{x^{l_*}}{x_*^{l_*+1}}-\frac{x_*^{l_*}}{x^{l_*+1}}\right)R(x)+I_1,
	\end{aligned}
\end{equation}
where
\begin{equation}
	I_1\equiv \int_0^\infty dx\,x^2\,R'(x)\left(\frac{x_*^{l_*}}{x^{l_*+1}}\right)R(x).
\end{equation}
For $l_*=1$,
\begin{equation}
	\begin{aligned}
		I_r
		=
		\int_0^{x_*} dx\,x^2\,R'(x)\left(\frac{x^{l_*}}{x_*^{l_*+1}}-\frac{x_*^{l_*}}{x^{l_*+1}}\right)R(x)+I_1+I_2,
	\end{aligned}
\end{equation}
where
\begin{equation}
	I_2\equiv -\int_0^\infty dx\,x^2\,R'(x)\left(\frac{x^{l_*}}{x_*^{l_*+1}}\right)R(x).
\end{equation}

\section{Ionization fluxes}\label{appendix_ionization}
In the nonrelativistic regime, the radiation of bosonic field initially within the cloud corresponds effectively to the bound-free transitions of GA. Here we adopt the ionization model established in Refs.~\cite{B4,Tomaselli:2023ysb}, the main result can be derived from an application of Fermi's golden rule to the tidally perturbed GA. The influences of bound-bound and free-free transitions on the ionization process are subleading and can be neglected \cite{B4,Tomaselli:2024bdd}.

Consider a state $\left|\psi(t)\right\rangle=c_{1}(t)\left|1\right\rangle+\int\frac{dk}{2\pi}c_{k}(t)\left|k\right\rangle$, where
$|1\rangle$ and $|k\rangle$ refer schematically to the bound and free eigenstate, with energy level $\omega_1$ and $\omega(k)$, respectively. Assuming the state is initially $|\psi(t=0)\rangle=|1\rangle$, under a periodic perturbation $\langle k|V_*(t)|1\rangle=\eta\,e^{-ig\Omega t}$ (with $\eta=\text{const}$, $g\in\mathbb{Z}^+$) it (in the approximation of first-order perturbation theory) evolves according to
\begin{equation}\label{single-mode_ionization}
\begin{aligned}
c_{k}(t)
&=-i\int_0^t dt'\langle k|V_*(t')|1\rangle \,e^{i[\omega(k)-\omega_n]t'}
\\
&=\eta \frac{1-e^{i[\omega(k)-\omega_1-g\Omega]t}}{\omega(k)-\omega_1-g\Omega}.
\end{aligned}
\end{equation}
In the limit $t\to \infty$, the transferred population is $|c_{k}|^2= 2|\eta|^2\frac{1-\cos (\Delta t)}{\Delta^2}\to 2\pi t |\eta|^2 \delta(\Delta)$, where $\Delta=\omega(k)-\omega_1-g\Omega$. Since $\omega(k)=\mu+\frac{k^2}{2\mu}$, the transition rate is
\begin{equation}
\begin{aligned}
-\frac{d|c_1|^2}{dt}
&=\int \frac{dk}{2\pi}\,\frac{d|c_k|^2}{dt}
\\
&=\int \frac{dk}{2\pi}\,\left[2\pi |\eta|^2 \delta(\Delta)\right]
=\left(\frac{\mu}{k}|\eta|^2\right)_{\Delta=0},\label{singe-mode_model}
\end{aligned}
\end{equation}
$\Delta=0$ gives the ionization condition $\mu+\frac{k^2}{2\mu}-\omega_1=g\Omega$.

Equation~\eqref{singe-mode_model} can be applied to the ionization of $|1\rangle$ in the more realistic situation. There is generally an infinite set of oscillation modes in the transition matrix element as given by Eqs.~\eqref{bound-free} and \eqref{rotation_of_Y}, which can be written as $\langle k|V_*|1 \rangle=\sum_{g=-\infty}^{\infty}\eta^{(g)}e^{-ig\Omega t}$. Note that different $l_*$ terms in the tidal potential can contribute to the same $g$-mode. The ionization is mediated by different $g$-modes toward different final states, Eq.~\eqref{single-mode_ionization} becomes
\begin{equation}
-\frac{d|c_1|^2}{dt}
=\sum_\text{final states}\left(\frac{\mu}{k}|\eta^{(g)}|^2\right)_{\Delta=0}.
\end{equation}
The associated radiation fluxes of cloud mass, energy and angular momentum (in the $z$-direction) are
\begin{align}
	P_\text{ion}&=(1+q)^{1/2}q^2\beta \alpha^5 \iota_1,
	\\
	-(\dot M_\text{c})_\text{ion}&=q^2\beta \alpha^3 \iota_2,
	\\
	\tau_{\text{ion},z}&=q^2\beta \alpha^2 M \iota_3,
\end{align}
where we introduced the nondimensional functions:
\begin{align}
	\iota_1&=x_a^{-3/2}\sum_\text{final states}\frac{g}{\tilde k}\frac{|\eta^{(g)}|^2}{q^2\alpha^2} \label{iota_1}
	,
	\\
	\iota_2&=\sum_\text{final states}\frac{1}{\tilde k}\frac{|\eta^{(g)}|^2}{q^2\alpha^2} \label{iota_2}
	,
	\\
	\iota_3&=\sum_\text{final states}\frac{m_*}{\tilde k}\frac{|\eta^{(g)}|^2}{q^2\alpha^2}. \label{iota_3}
\end{align}
For a generic elliptical orbit and a generic GA state, the $\iota$ functions could depend on $\{q,x_a,e,i,\varphi_0\}$ (due to the cylindrical symmetry of the problem, they should be independent of $\phi_0$). The ionization flux of vector GA can be computed directly using the angular integral \eqref{I_Omega}, but since the Cartesian components of the NR wave function of vector GA can always be decomposed into a superposition of scalar GA eigenstates and with the net ionization flux given by the sum over the flux of each component, we need only to compute the ionization fluxes of the relevant scalar GA states.

The above analysis is performed in the $x'y'z'$ frame (since we consider a fixed Keplerian orbit, $\phi_0$ can be set to zero, then $x'y'z'=xyz$, see Fig.~\ref{orbit_2}), a convenient way to obtain the full ionization torque is to work in the $XYZ$ frame\footnote{The computations in the two frames are equivalent. We confirmed this with a computation of the ionization torque for an inclined circular orbit in the scalar $|211\rangle$ cloud, the ionization power $P=\Omega\tau_Z$ computed in the $XYZ$ frame agrees with that computed in the $x'y'z'$ frame.} \cite{Tomaselli:2023ysb}, where the companion travels on the $XY$ plane (the equatorial plane in this frame), simplifying the computation of level mixings, but with the price of treating a nonspherical ($l\ge 1$) scalar GA eigenstate as a superposition of ``hyperfine'' levels (which differ from each other only by the azimuthal quantum number, but are unrelated to the hyperfine structure of GA) according to Eq.~\eqref{rotation_of_Y}. In this frame $\tau_Z$ (angular momentum flux in the $Z$-direction) can be computed similarly as for $\tau_z$ in the $x'y'z'$ frame. The angular momentum flux in the $X$-direction can be obtained from $\tau_z=\boldsymbol{\tau}\cdot\mathbf{e}_z=(\tau_X\mathbf{e}_X+\tau_Z\mathbf{e}_Z)\cdot\mathbf{e}_z$ as $\tau_X=(\tau_z-\tau_Z\cos i)/\sin i$. For a nonspherical state, the ionization fluxes for an inclined eccentric orbit would be $\varphi_0$-dependent, since more than one bound state in the $XYZ$ frame can be ionized into the same free state, leading to interference in the transition matrix elements.

\section{Cloud depletion effect}\label{appendix_DC}
In Sec.~\ref{(ii)} we assumed that the (time-averaged) metric perturbation sourced by the cloud is stationary, this is the case if the cloud is dominated by a single eigenstate and its amplitude does not evolve. Recall that the mass density of the cloud is given by $\rho(t,\mathbf{r})=\beta(t=0) M\sum_i|\Psi_i(t)|^2$, if we choose the normalization $\int d^3r\,\sum_i|\Psi_i(t=0,\mathbf{r})|^2=1$. E.g., if the cloud is occupied by multiple eigenstates, the density and the Newtonian potential of the cloud would undergo temporal oscillations at frequency given by the energy level difference of the involved states, which could be non-negligible if the oscillation is sufficiently slow. This is exactly the origin of backreaction from the companion-induced GA transitions.

Even for a single occupied eigenstate, the Newtonian potential $\Phi\propto M_\text{c}$ also undergoes temporal evolution if we take into account the depletion of cloud (DC) due to its intrinsic GW emission (with mass depletion rate $P_\text{gw,c}$), and in a binary system, the GA is further ionized by its companion. Combing these two effects we have $\dot M_\text{c}=-P_\text{gw,c}+(\dot M_\text{c})_\text{ion}$. If we neglect the linear momentum loss of GA in its rest frame due to the GW radiation of the cloud (mass loss of this type is called isotropic \cite{hadjidemetriou1963two} or the Jeans' mode \cite{1963ApJ...138..471H}) as well as the additional backreaction of the ionization process, the orbital dynamics (without considering the dissipative effects) is still given by $\ddot{\mathbf{r}}_*=-\frac{M_\text{tot}}{r^3_*}\mathbf{r}_*-\nabla\Phi$, only with $\Phi(t,\mathbf{r})$ being \textit{slowly} time-varying at given position $\mathbf{r}$.

The gravitational acceleration can be split into the radial and nonradial parts:
\begin{equation}
	-\nabla\Phi(t,r,\theta)= -\frac{\tilde M_\text{c}(t,\mathbf{r}_*(t))}{r^2}\mathbf{e}_r+(-\nabla\Phi\cdot\mathbf{e}_\theta)\,\mathbf{e}_\theta,
\end{equation}
where we define an effective cloud mass:
\begin{equation}
	\tilde M_\text{c}(t,\mathbf{r})\equiv r^2\nabla\Phi\cdot\mathbf{e}_r.
\end{equation}
This corrects the treatment in Ref.~\cite{Cao:2023fyv}, where $\tilde M_\text{c}$ was taken to be $-r\Phi$. Then we consider the osculating orbit defined by the total mass $M+\tilde M_\text{c}$, instead of $M$. Since the evolution of $\Phi$ is slow, this time-dependence essentially makes no difference to the nonradial acceleration, the time-dependent $\tilde M_\text{c}$ however gives rise to an effective acceleration \cite{hadjidemetriou1963two}:
\begin{equation}
	\mathbf{F}_\text{DC}=-\frac{1}{2}\frac{\partial_t\tilde M_\text{c}+\mathbf{v}\cdot\nabla \tilde M_\text{c}}{M_\text{tot}+\tilde M_\text{c}}\mathbf{v}
	\approx
	-\frac{1}{2}\frac{\partial_t\tilde M_\text{c}}{M_\text{tot}}\mathbf{v},
\end{equation}
where we kept only the term stemming from the local time-dependence of $\tilde M_\text{c}$. In the case of $M_\text{c}$ being static, the other term $\mathbf{F}\approx-\frac{\mathbf{v}\cdot\nabla \tilde M_\text{c}}{2 M_\text{tot}}\mathbf{v}$ indeed gives the same precession rate as $\mathbf{F}=-\nabla \Phi$ in the limit $M+\tilde M_\text{c} \approx  M$, the two approaches are thus consistent. Here we also neglected the possible baryonic accretion of the central BH, which would increase $M_\text{tot}$ and thus compete with the cloud depletion process. The temporal evolution of the effective cloud mass is given by $	\partial_t\tilde M_\text{c}=\tilde\beta\,\dot M_\text{c}$, with $\tilde\beta(\mathbf{r})\equiv\tilde M_\text{c}/M_\text{c}$. $\mathbf{F}_\text{DC}$ above gives the evolution of orbital elements, such as
\begin{equation}
	\frac{\langle\dot a\rangle_\text{DC}}{a}\approx-\frac{a}{M_\text{tot}^2}\left\langle v^2\partial_t\tilde M_\text{c}\right\rangle.
\end{equation}
Correspondingly the evolution of $\Omega$ is given by\footnote{In Ref.~\cite{Cao:2023fyv} the first term on the RHS was incorrectly neglected, thus underestimating the cloud depletion effect on the frequency evolution, e.g., by a factor of $4/3$ when $\tilde M_\text{c}$ is essentially constant within one orbital period, the associated braking index is $1/2$ rather than unity.}
\begin{equation}
	\frac{\langle\dot\Omega\rangle}{\Omega}
	=
	\frac{1}{\Omega}\left\langle\frac{d}{dt}\sqrt{\frac{M_\text{tot}+\tilde M_\text{c}}{a^3}}\right\rangle
	\approx
	\frac{1}{2}\frac{\langle\partial_t\tilde M_\text{c}\rangle}{M_\text{tot}}-\frac{3}{2}\frac{\langle \dot a \rangle}{a}
	.
\end{equation}

For the $|1011\rangle$ cloud, $\tilde\beta(x)=1+e^{-2 x} \left(-2 x^2-2 x-1\right)$, if $(\dot M_\text{c})_\text{ion}$ is neglected, we obtain the following results:
\begin{widetext}
	\begin{align}
		\langle\tilde \beta\rangle_\text{DC}
		&=1+
		e^{-2 x_a} \Big[\left(2 e^2 x_a^2+6 x_a^2+7 x_a+3 \right)e\, I_1(2 e x_a)
		-\left(6 e^2 x_a^2+3 e^2 x_a+2 x_a^2+2 x_a+1\right) I_0(2 e x_a)\Big]
		,
		\\
		\frac{\left\langle \dot a/a \right\rangle_\text{DC}}{P_\text{gw,c}/M_\text{tot}} &=1+
		e^{-2 x_a} \Big[\left(-2 e^2 x_a^2+2 x_a^2- x_a-3 \right)e\, I_1(2 e x_a)
		+\left(2 e^2 x_a^2+3 e^2 x_a-2 x_a^2-2 x_a-1\right) I_0(2 e x_a)\Big]
		,
	\end{align}
\end{widetext}
where $I_n(z)$ is the modified Bessel function of the first kind. Since the ionization flux provides a secular description for the backreaction from ionization, it would be inaccurate to treat $(\dot M_\text{c})_\text{ion}$ as constant within one orbital period, but this problem is actually unimportant. As can be checked, $(\dot M_\text{c})_\text{ion}$ becomes non-negligible (and finally dominant) when the orbital radius is sufficiently small, but in such case $P_\text{DC}\equiv \frac{qM\alpha^2}{2x_a^2}\langle \dot x_a\rangle_\text{DC}$ also becomes negligible relative to $P_\text{ion}$.

\begin{figure}[htb]
	\centering
	\includegraphics[width=0.47\textwidth]{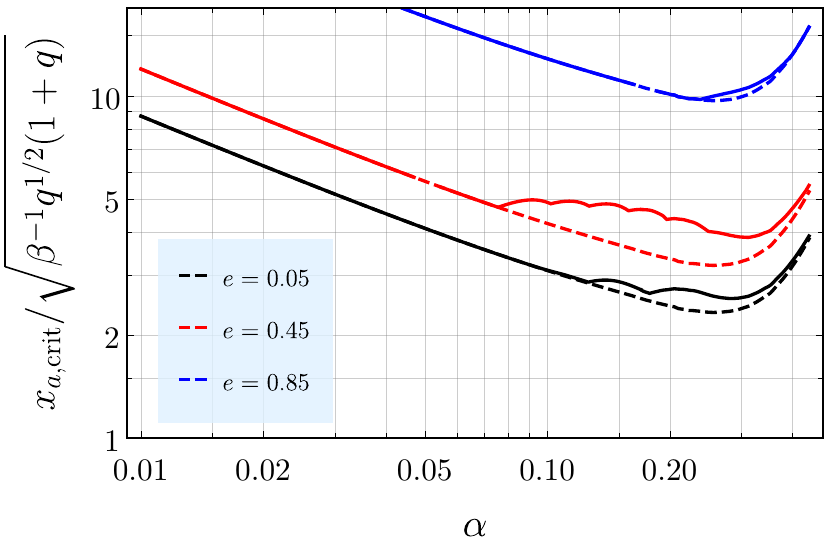}
	\caption{Critical semimajor axis of outspiral in a saturated $|1011\rangle$ cloud for $\beta/\alpha=0.1$ and $q=10^{-3}$. Dashed lines correspond to the estimation \eqref{X_crit}.}\label{x_crit}
\end{figure}

Since $\dot M_\text{c}<0$, $\mathbf{F}_\text{DC}$ tends to increase the orbital separation, and it becomes more significant at larger radius. When $\alpha$ is moderately small, the critical semimajor axis of outspiral (above which $\langle\dot x_a\rangle>0$) is approximately given by the balance between $P_\text{DC}$ and $P_\text{gw}$ alone (with $\dot M_\text{c}\approx -P_\text{gw,c}$), even for a large orbital eccentricity. Together with $\langle \tilde\beta\rangle_\text{DC}\approx 1$ this leads to the estimation
\begin{equation}
	x_{a,\text{crit}}\approx\left[\frac{64}{5}\frac{q(1+q)^2\alpha^8}{p(\alpha)\,\beta^2}f(e)\right]^{1/4}, \label{X_crit}
\end{equation}
as depicted in Fig.~\ref{x_crit}. In the presence of extra dissipation, $x_{a,\text{crit}}$ is further enlarged, so this gives a lower bound. Taking $x_{a,\text{crit}}$ into account imposes upper bounds on the cloud mass before the low-frequency fine or hyperfine transitions during an inspiral \cite{Cao:2023fyv}. The cloud depletion effect due to $P_\text{gw,c}$ can possibly be probed in certain GA-pulsar binaries through long-term timing observations \cite{Cao:2023fyv}. But for the inspiraling systems with $M\gtrsim 10^4M_\odot$ , even in the case of vector $|1011\rangle$ cloud, the orbital frequency at which the cloud depletion matters would be well below the LISA band if a reasonably small value of $\alpha$ is assumed, and the cloud depletion effect can be safely neglected for $x\lesssim 1$.

\section{Companion-induced energy level correction, Bohr resonances and nonresonant transitions}\label{appendix_Bohr}
Here we briefly discuss the consequences of bound state mixings for an inspiral in the $|1011\rangle$ cloud. Neglecting the free states and the cloud's self-gravity, a bound state $|\psi(t)\rangle=\sum_i c_i(t)|i\rangle$ evolves according to the Shr\" {o}dinger equation: $i\dot c_i=H_{ij}c_j$ with $H_{ij}\equiv\langle i|V_*|j \rangle$. Since the energy levels are generally complex, $H_{ij}$ is non-Hermitian. Using Eqs.~\eqref{bound-state-transition} and \eqref{rotation_of_Y}, the transition matrix element is given by
\begin{equation}
\langle i'|V_*|i\rangle
=
\sum_{g=-\infty}^{\infty}\eta^{(g)}e^{-ig\Omega t}e^{i(\omega'-\omega)t},
\end{equation}
with the Fourier coefficients:
\begin{equation}
\begin{aligned}
\eta^{(g)}=-\sum_{l_*,m_b}\frac{4\pi q\alpha I_\Omega}{2l_*+1}
d^{l_*}_{m_*,m_b}(-i)\,Y_{l_*m_b}\left(\frac{\pi}{2},0\right)\,C_{l_*,m_b}^{(g)},
\end{aligned}
\end{equation}
and
\begin{equation}
C_{l_*,m_b}^{(g)}(x_a,e) = \frac{\Omega}{2\pi}\int_0^{2\pi/\Omega}dt\,\left[\frac{I_r(l_*)}{r_\text{c}}\,e^{-im_b(\varphi+\varphi_0)}\right] \,e^{ig\Omega t}.
\end{equation}
Apart from the level mixings, $\langle i|V_*|i\rangle$ gives a correction to the energy level of $|i\rangle$. For the $|1011\rangle$ state, this correction comes from $l_*=m_*=0$ term in the tidal potential, thus $\langle 1011|V_*|1011\rangle=\sum_ g \eta_0^{(g)}e^{-ig\Omega t}$, with $\eta_0^{(g)}(t)=-q\mu\alpha^2\langle I_r e^{ig\Omega t}\rangle$. This correction is negligible compared with the unperturbed level $\omega^{(1011)}\approx \mu-\mu\alpha^2/2$ if $q\ll 1$, but its magnitude can be larger than the self-gravity correction\footnote{See Refs.~\cite{Siemonsen:2022yyf,may2024selfgravityeffectsultralightboson} for recent investigations in the relativistic regime.} \cite{Baryakhtar:2017ngi} $\Delta \omega^{(1011)}_{\text{self}, \Phi}=\langle 1011|\mu\Phi|1011\rangle=-\frac{5}{8}\beta\mu\alpha^2$ if $q>\beta$. The $g\ne 0$ terms correspond to oscillating modulations, which are strongly suppressed relative to the static correction $\eta_0^{(0)}$ even for a high eccentricity.

For the cloud initially occupied by a single bound eigenstate $|1\rangle$, resonant transition from $|1\rangle$ to the bound state $|2\rangle$ occurs when $\omega_R^{(2)}-\omega_R^{(1)}-g\Omega=0$. Only a single harmonic number $g\in \mathbb{Z}$ is involved in this process, but it can happen at multiple frequencies $\Omega_*^{(g)}=[\omega^{(2)}_R-\omega^{(1)}_R]/g$ if the orbit is not circular, the highest possible one being $\Omega_*^{(1)}=|\omega^{(2)}_R-\omega^{(1)}_R|$. For the vector GA, a transition can have multiple final states \cite{B3} with energy level difference much smaller than that of transition; the Bohr resonances with multiple states can also occur simultaneously in the generic case of inclined eccentric orbit. Here for simplicity we discuss only the situation of two-state resonance. Neglecting the diagonal and self-gravity corrections, the relevant equation of motion is
\begin{equation}\label{two-mode}
\begin{aligned}
i\dot C_1 &=\omega^{(1)}C_1+\left[\eta^{(g)}e^{-ig\Omega t}\right]^*C_2,
\\
i\dot C_2
&=\omega^{(2)}C_2+\eta^{(g)}e^{-i g\Omega t}C_1,
\end{aligned}
\end{equation}
where we introduced $c_i\equiv e^{i\omega^{(i)}t}C_i$, such that $|\psi(t)\rangle=\sum_i C_i(t)\,\left[e^{i\omega^{(i)}t}|i\rangle\right]$. Under the unitary transformation $U=\text{diag}(e^{ig\Omega t/2}, e^{-ig\Omega t/2})$, the Hamiltonian $H=\left(\begin{smallmatrix}
\omega^{(1)} & [\eta^{(g)}]^* e^{ig\Omega t}
\\
\eta^{(g)} e^{-ig\Omega t} & \omega^{(2)}
\end{smallmatrix}\right)$ for $\{C_i\}$ becomes $H'=U^\dagger HU-iU^\dagger \dot U=\left(\begin{smallmatrix}
\omega^{(1)}+g\Omega/2 & [\eta^{(g)}]^*
\\
\eta^{(g)} & \omega^{(2)}-g\Omega/2
\end{smallmatrix}\right)$. If the backreaction of resonance is non-negligible, since the level mixing depends on $\{x_a,e,i\}$, Eq.~\eqref{two-mode} is coupled with Eqs.~\eqref{dissip_i}, \eqref{dissip_phi0}, \eqref{dissip_e} and \eqref{dissip_a}, as explored in Refs.~\cite{Takahashi:2021yhy,Tomaselli:2024bdd,Boskovic:2024fga} for the scalar GA. Otherwise, we can use the approximation  $\Omega(t)\approx\Omega_*^{(g)}+\dot\Omega t$, with the adiabaticity of the transition governed by the Landau-Zener (LZ) parameter \cite{B3} $z^{(g)}\equiv 2|\eta^{(g)}|^2/|g\dot\Omega|$. If $z\ll 1$, the transition would be rather nonadiabatic, thus with very limited population being transferred\footnote{In the LZ approximation, the asymptotic final population $|C|^2$ of an initially unoccupied state is $1-e^{-\pi z}$.}. Although such a resonance event itself may have interesting observational signatures, its occurrence is not expected to change the state of the cloud. In contrast, the floating adiabatic resonance features non-negligible backreaction and could leave considerable imprints on both the cloud and the orbit \cite{Tomaselli:2024bdd}.

In the case of $|1011\rangle$ state, the selection rule for its mixing with $|n'l'j'm'\rangle$ state from the angular integral \eqref{I_Omega} reads \cite{B3} $|m_*=m'-1|\le l_*=l'\in[|j'-1|,j'+1]$. Note that generically, $|n,0,j,1\rangle $ does not mix with $|n',0,j',-1\text{ or } 0\rangle$ and $|n',l',j',-l'\text{ or }-l'-1\rangle$; so $|1011\rangle$ does not mix with these states, in particular $|101,-1\rangle$ and $|1010\rangle$, this means a cloud in the $|1011\rangle$ state can only experience sinking Bohr resonances with $|n'\ge2,l',j',m'\rangle$ at orbital frequency $\Omega_*^{(g)}\approx\frac{\mu \alpha^2}{2g}\left[1-(n')^{-2}\right]$, corresponding to the radius 
\begin{equation}
x_*^{(g)}\approx(1+q)^{1/3}\left[\frac{2g}{1-(n')^{-2}}\right]^{2/3}.
\end{equation}
E.g., for the equatorial orbit, there can be resonances with $|n'011\rangle$\footnote{Note the self-gravity of $|1011\rangle$ cloud also leads to its mixings with $|n'011\rangle$.} (through $l_*=0$), $|2122\rangle$ and $|21j'0\rangle$ (through $l_*=1$), $|3233\rangle$ and $|32j',-1\rangle$ (through $l_*=2$), and states with higher $n'$. The nonadiabaticity of such sinking Bohr transition will be enhanced by the backreaction of ionization and the transition itself. As a concrete example, we list the LZ parameters of the last two possible ($g=1,2$) resonances with $|3233\rangle$ (only $l_*=|m_*|=2$ contributes to this resonance) for several choices of $\{i,e\}$ in Table~\ref{LZ_table}, using $|\dot\Omega|=[3(1+q)^{1/2}q^{-1}x_a^{-1/2}\alpha (P_\text{gw}+P_\text{ion})M^{-2}]_{x_a=x_*^{(g)}}$. Note that $|3233\rangle$ is a growing state when the SR instability of $|1011\rangle$ saturates, hence the transferred population will not be absorbed by the host BH (different from the normal situation in scalar GA), but the growth rate is typically much smaller than $|\eta^{(g)}|$. For the parameters used in Table.~\ref{LZ_table}, the orbital period decay around resonance in all cases is controlled by ionization and all the resonances are nonadiabatic. Note however when $P\approx P_\text{ion}$ and $q\ll 1$, $z$ is proportional to $q/\beta$ and is independent of $\alpha$, thus for sufficiently small cloud or large companion mass, the transition can be adiabatic. We leave this possibility for future study.

\begin{table}[htb!]
	\renewcommand{\arraystretch}{1}
	\begin{center}
		\begin{tabular}{ccccccc}
			\hline
			\hline
			$\quad i \quad$ & $\quad e \quad$ & $\quad z^{(1)} (x_*\approx1.71) \quad$ & $\quad z^{(2)} (x_*\approx2.72) \quad$ \\
			\hline
			$0$& $0.01$ & $8.9\times 10^{-10}$ & $3.0\times 10^{-5}$ \\
			$\pi/3$& $0.01$ & $2.2\times 10^{-10}$ & $9.5\times 10^{-6}$ \\
			$0$& $0.3$ & $6.2\times 10^{-7}$ & $2.3\times 10^{-6}$\\
			$0$& $0.7$ & $4.5\times 10^{-7}$ & $8.9\times 10^{-8}$\\
			\hline
			\hline
		\end{tabular}
	\end{center}
	\caption{LZ parameters of $|1011\rangle \leftrightarrow |3233\rangle$ resonance for $\alpha=\beta=0.05$, $q=10^{-4}$ and $\varphi_0=0$.}
	\label{LZ_table}
\end{table}

\begin{figure*}[t]
	\centering
	\includegraphics[width=0.47\textwidth]{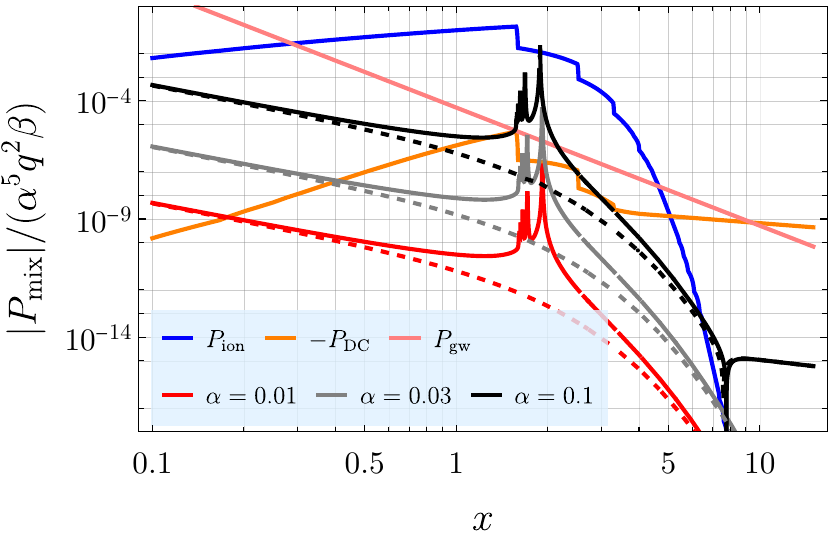}
	\quad
	\includegraphics[width=0.47\textwidth]{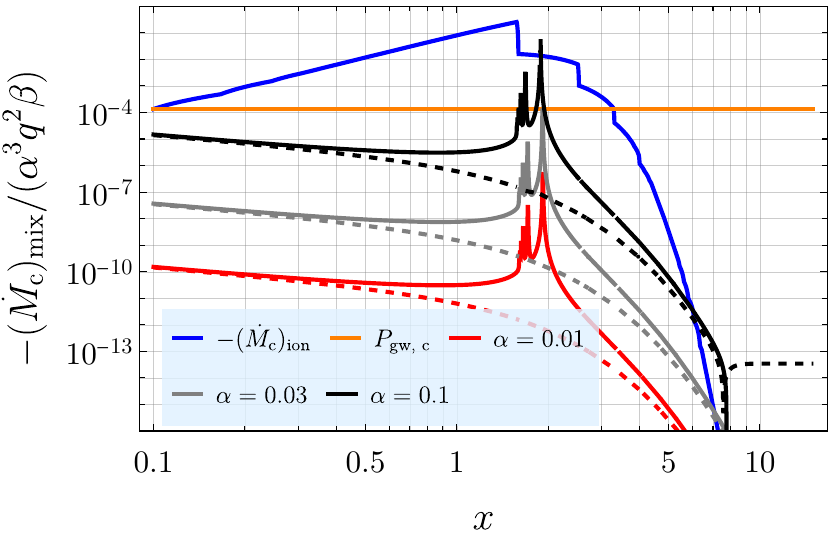}
	\caption{Comparison between $P_\text{ion}$, $P_\text{DC}$, $P_\text{gw}$, $P_\text{mix}$ (left panel) and between $-(\dot M_\text{c})_\text{ion}$, $P_\text{gw,c}$, $-(\dot M_\text{c})_\text{mix}$ (right panel) as given by the estimation \eqref{P_mix} for equatorial circular orbit in a saturated vector $|1011\rangle$ cloud with $\omega_{I}^{(1011)}(\chi)=0$. Solid (dashed) line corresponds to the counterrotating (corotating) orbit. The sharp peaks appear when the resonance condition is met. The parameters used for $P_\text{gw}$, $P_\text{DC}$, $(\dot M_\text{c})_\text{ion}$ and $P_\text{gw,c}$ are $\alpha=10\beta=0.03$, $q=10^{-4}$.}\label{1011_P_mix}
\end{figure*}

Outside the resonance band, bound state mixings may also affect the orbital evolution, especially when the mixed state has a large decay rate \cite{WY_3}. In the case of equatorial corotating ($i=0$) circular orbit, this effect can be estimated as follows (for the counterrotating orbit ($i=\pi$), $\Omega$ is replaced by $-\Omega$). Under the transformation $U_{ij}=\delta_{ij}\,e^{i[\omega^{(i)}-m_i\Omega]t}$, the bound-state Hamiltonian $H_{ij}$ for $\{C_i\}_i$ becomes
\begin{equation}
H_{ij}'=(U^{-1} HU-iU^{-1} \dot U)_{ij}=[\omega^{(i)}-m_i\Omega]\delta_{ij}+\mathcal{H}_{ij},
\end{equation}
with $\mathcal{H}_{ij}\equiv H_{ij}e^{i\{(m_i-m_j)\Omega -[\omega^{(i)}-\omega^{(j)}]\}t}$ being approximately static due to the adiabaticity of orbital evolution. The energy level correction in the second order nondegenerate perturbation theory is given by
\begin{equation}
	\Delta\omega^{(i)}= \mathcal{H}_{ii}+\sum_{j\ne i} \frac{|\mathcal{H}_{ij}|^2}{\omega^{(i)}-\omega^{(j)}-(m_i-m_j)\Omega}.\label{correction}
\end{equation}
For the $|1011\rangle$ cloud, the dominant contribution to $\Delta\omega_I^{(1011)}$ comes from the transitions to the $m_j=0$ states, whose decaying rates ($-\omega_I^{(j)}>0$) are typically large compared with their mixings with $|1011\rangle$. The evolution of GA due to the bound state mixings is also expected to be dominated by such transitions \cite{WY_3}. The change of the angular momentum of an initially saturated $|1011\rangle$ cloud due to the transitions is then approximately given by $(\dot S_\text{c}+\dot J)/(M_\text{c}/\mu)=\frac{d |C^{(1011)}|^2}{dt}+\sum_j [m_j\frac{d |C^{(j)}|^2}{dt}-2\omega_{I}^{(j)}|C_j|^2m_j]\approx \frac{d |C^{(1011)}|^2 }{dt}\approx 2\Delta\omega_I^{(1011)}|C^{(1011)}|^2\approx 2\Delta\omega_I^{(1011)}$. According to Eq.~\eqref{dot_L}, the effective dissipation power of cloud mass and orbital energy would be
\begin{equation}
\begin{aligned}\label{P_mix}
	(\dot M_\text{c})_\text{mix}&\approx M_\text{c}\frac{d |C^{(1011)}|^2 }{dt}\approx2\beta M \Delta \omega_I^\text{(1011)},
	\\
	P_\text{mix}&\approx\pm \frac{\alpha^3}{x^{3/2}}\frac{\dot S_\text{c}}{M}\approx\pm 2m_1\beta\frac{\alpha^3}{x^{3/2}} \frac{\Delta\omega_I^{(1011)}}{\mu},
\end{aligned}
\end{equation}
where $+/-$ stands for corotating/counterrotating orbit. The resonance events can be more appropriately treated individually, so the more interesting aspect of this estimation is the prediction for orbital frequency away from the ``poles'' $\Omega=[\omega_{R}^{(i)}-\omega_{R}^{(j)}]/(m_i-m_j)$. We show the results for $\alpha=0.01$, 0.03, 0.1 in Fig.~\ref{1011_P_mix}. In this computation we use the analytical spectrum of vector GA derived in Ref.~\cite{B2} (neglecting the self-gravity corrections), and the summation in Eq.~\eqref{correction} is performed over all modes with $m'\in[-5,5]$ and $n'-l'-1\in[0,5]$ to ensure convergence. For a moderately small $\alpha$, $P_\text{mix}$ and $(\dot M_\text{c})_\text{mix}$ appear to be negligible compared with other effects.

\begin{figure*}[htb!]
	\centering
	\includegraphics[width=0.46\textwidth]{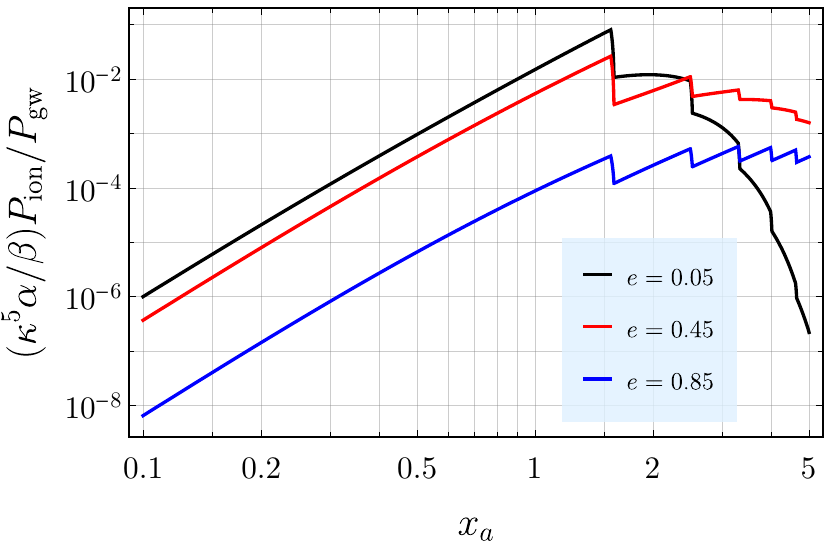}
	\quad
	\includegraphics[width=0.47\textwidth]{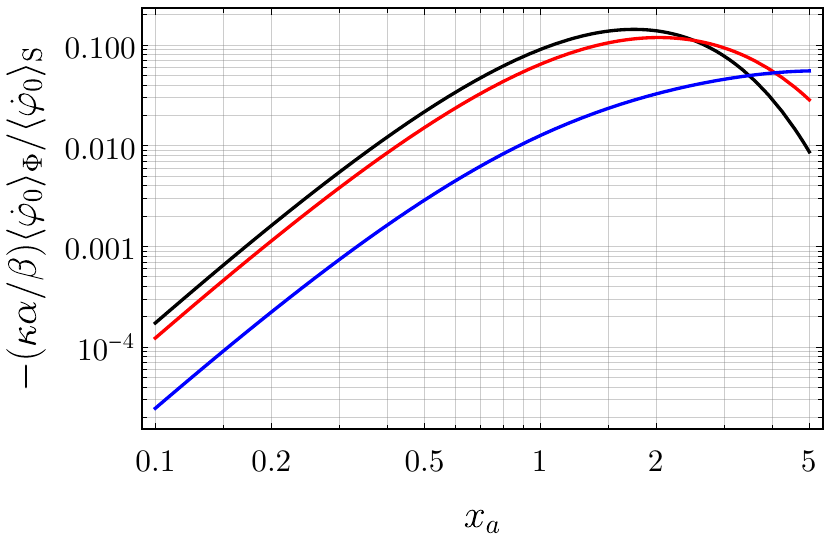}
	\caption{$P_\text{ion}/P_\text{gw}$ and $\langle \dot\varphi_0\rangle_{\Phi}/\langle \dot\varphi_0\rangle_\text{S}$ for $q\ll 1$.}\label{given-frequency}
\end{figure*}

\section{Orbital evolution at given frequency}\label{appendix_given-frequency}
Given the host BH mass $M$ and the orbital frequency $\Omega$, $\alpha=\kappa\sqrt{x_a}$, with $\kappa\equiv (M\Omega/\sqrt{1+q})^{1/3}$. In Fig.~\ref{given-frequency} we plot the ratios $P_\text{ion}/P_\text{gw}$ and $\langle \dot\varphi_0\rangle_{\Phi}/\langle \dot\varphi_0\rangle_\text{S}$ as functions of $x_a$ for given $\kappa$ and $\beta/\alpha$ in the small $q$ limit. $x_a$ is related to the boson mass via $\mu=\kappa\sqrt{x_a}/M$. It is seen that for $x_a\lesssim 1$, both the ionization and the cloud-induced periastron precession become more significant relative to the vacuum effects for a larger boson mass. For $x_a>x^{(2)}\approx 2.5$, the ratio $P_\text{ion}/P_\text{gw}$ is nonmonotonic with respect to $\alpha$ unless the eccentricity is small.

\newpage
\bibliography{paper}
\end{document}